\documentclass[structabstract]{aa}  
\usepackage{graphicx}
\usepackage{txfonts}

\usepackage{url}
\usepackage{aalongtable}
\usepackage{enumerate}

\usepackage{natbib}
\bibpunct{(}{)}{;}{a}{}{,} % to follow the A&A style

\usepackage{amsmath}

\begin{document}
\title{Mining the UKIDSS GPS: star formation and embedded clusters
  \thanks{Appendices A, B and C are only available in electronic form via http://www.edpsciences.org}}

\author{O. Solin\inst{1,2},
        E. Ukkonen\inst{1}
        \and
        L. Haikala\inst{3,2}
        }

\institute{University of Helsinki, Department of Computer Science\\
           P.O. Box 68, FI-00014 University of Helsinki, Finland\\
\email{otto.solin@helsinki.fi}
           \and
           University of Helsinki, Department of Physics, Division of Geophysics and Astronomy\\
           P.O. Box 64, FI-00014 University of Helsinki, Finland
           \and
           Finnish Centre for Astronomy with ESO\\
           University of Turku, V\"ais\"al\"antie 20, FI-21500 PIIKKI\"O, Finland
           }

\date{}

\abstract
{Data mining techniques must be developed and applied to analyse the large public data bases containing hundreds to thousands of millions entries.}
{To develop methods for locating previously unknown stellar clusters from the UKIDSS Galactic Plane Survey catalogue data.}
{The cluster candidates are computationally searched from pre-filtered catalogue data using a method that fits a mixture model of Gaussian densities and background noise using the Expectation Maximization algorithm. The catalogue data contains a significant number of false sources clustered around bright stars. A large fraction of these artefacts were automatically filtered out before or during the cluster search. The UKIDSS data reduction pipeline tends to classify marginally resolved stellar pairs and objects seen against variable surface brightness as extended objects (or "galaxies" in the archive parlance). 10\% or $66 \times 10^6$ of the sources in the UKIDSS GPS catalogue brighter than $17^\mathrm{m}$ in the \textit{K} band are classified as "galaxies". Young embedded clusters create variable NIR surface brightness because the gas/dust clouds in which they were formed scatters the light from the cluster members. Such clusters appear therefore as clusters of "galaxies" in the catalogue and can be found using only a subset of the catalogue data. The detected "galaxy clusters" were finally screened visually to eliminate the remaining false detections due to data artefacts. Besides the embedded clusters the search also located locations of non clustered embedded star formation.}
{The search covered an area of 1302 deg$^2$ and 137 previously unknown cluster candidates and 30 previously unknown sites of star formation were found.}
{}

\keywords{open clusters and associations: general -- methods: statistical -- catalogs -- surveys -- infrared: stars}
\maketitle
%
%________________________________________________________________

\section{Introduction}

Several large digital data archives have become publicly available during the last decade. The archive data of stars and extragalactic objects has been extracted from dedicated large imaging surveys in wavelengths from optical (e.g. SDSS) through near-infrared (NIR) (e.g. the Two Micron All Sky Survey (2MASS; \citet{Skrutskie}), the UKIRT Infrared Deep Sky Survey (UKIDSS; \citet{UKIDSS}) to mid-infrared (MIR) (e.g. GLIMPSE) and far-infrared (FIR) (e.g. MIPSGAL). The catalogues contain hundreds of millions (e.g. SDSS) to thousands of millions (e.g. UKIDSS when finished) objects. Extracting information from a survey containing terabytes of data can naturally be done in the traditional way, case by case, in small restricted areas. But to really optimise the use of all the data, data mining techniques have to be applied. Data mining will allow to identify 'hidden' patterns and relations, which are not obvious, within the data \citep{DMintro}. In astronomy data mining methods have been applied to various research areas such as object classification, forecasting sunspots, and selection of quasar candidates \citep{sabine}.

The major part of star formation, be it low- or high-mass stars, takes place in clusters. The clusters are not bound and will eventually disrupt e.g. because of the Galactic differential rotation \citep{Blaauw}. The stellar clusters trace therefore the recent Galactic star formation. The younger the clusters are the more compact they are and the more closely they are associated with the interstellar gas and dust clouds they formed in. Detailed study of young clusters still associated with their parent cloud will provide information on the star formation process and the stellar initial mass function (IMF).

At the moment some 2000 Galactic stellar clusters are known. This is only a small fraction of the estimated total population of which a major part is obscured by interstellar dust to us and can not be observed in optical wavelengths. However, the extinction decreases at longer wavelengths and already at 2.2 microns in the NIR the extinction in magnitudes is only 11 percent of that in the \textit{V} band \citep[e.g.][]{Rieke}. The ongoing UKIDSS Galactic plane survey (GPS) is three magnitudes deeper than 2MASS and offers the possibility of detecting stellar clusters which are either more distant and/or more extincted than those visible in 2MASS. The UKIDSS GPS will cover 1800 deg$^2$ of the northern Milky Way in \textit{JHK} to the limiting magnitudes of J=20\fm0,H=19\fm1,K=18\fm1. The survey began in May 2005 and when finished will provide an estimated $\sim 1-2 \times 10^9$ detections (mainly stellar sources) in three passbands, i.e. $\sim 3$ times that of the 2MASS whole sky survey. Searching automatically for a stellar cluster in the complete UKIDSS GPS is possible only using data mining techniques.

Clusters from infrared archive data have been searched for by \citet{Dutra} and \citet{BDBarbuy,BDSoares} by visual inspection of 2MASS images. \citet{MCM} searched the GLIMPSE mid-infrared survey for clusters using an automated algorithm and visual inspection of images. \citet{FSR} (FSR) used 2MASS star density maps to locate clusters. \citet{SamuelLucas} and \citet{LucasPres08,LucasList,LucasPres11} applied the ideas from \citet{MCM} to look for clusters from the the UKIDSS GPS. In a recent paper \citet{oldClsFSR} applied the code by \citet{SamuelLucas} developed for the cluster search in UKIDSS GPS data to investigate the old star clusters in the FSR list \citep{FSR}. Many of these newly detected clusters or cluster candidates have not yet been studied in detail.

This paper presents an application of Gaussian mixture modelling, optimised with the Expectation Maximization (EM) algorithm \citep{EM} to automatically locate stellar clusters in the UKIDSS GPS. The search algorithm and filtering of the catalogue artefacts have been described in detail. The search has so far been applied to the UKIDSS GPS DR7 covering an area of 1302 deg$^2$. The data is described in Sect. \ref{sec:theData} and the search method and results in Sects. \ref{sec:searchMethod} and \ref{sec:results}. In Sect. \ref{sec:discussion} the data mining approach to cluster search, the results, supplementary information on the cluster candidates and selected individual cluster candidates are discussed. Conclusions are drawn in Sect. \ref{sec:conclusions}.

%__________________________________________________________________

\section{The data}\label{sec:theData}
UKIDSS is conducted with the Wide Field Camera (WFCAM; \citet{Casali}) mounted on the United Kingdom Infrared Telescope (UKIRT) on Mauna Kea. WFCAM consists of four 2048x2048 Rockwell devices and a single exposure covers an area of 0.21 deg$^2$. The photometric system used by UKIDSS is described in \citet{Hodgkin} and \citet{Hewett}. The WFCAM Science Archive \citep[WSA; Irwin et al. in preparation;][]{Hambly} holds the UKIDSS image and catalogue data \mbox{products}. The catalogue data is used for the automated search, and the image data for visual inspection of the cluster candidate areas given by the detection algorithm. The WSA releases the data in stages. The current 7th release for GPS covers 1302 deg$^2$ for the UKIDSS GPS \textit{K} filter. Of this 819 deg$^2$ are covered in the \textit{J} and \textit{H} filters. This study uses all the data covered in the \textit{K} band. Stars brighter than $K=10^\mathrm{m}$ from the 2MASS survey are used for locating potential false positive clusters (see Figs. \ref{diffrSpikes} and \ref{brightAtBorder}).

%__________________________________________________________________

\section{Search method}\label{sec:searchMethod}
The search method takes pre-filtered catalogue data, divided into overlapping bins of size 4\arcmin\ by 4\arcmin, and performs a maximum likelihood fitting of a mixture of a Gaussian density and a uniform background. On each bin the fitting is done using the standard Expectation Maximization (EM) algorithm that is widely applied in a variety of sciences, and generally for data clustering in machine learning. The EM-algorithm has been applied for clustering in astronomy by \citet{MCM} to discover new star clusters in the GLIMPSE survey, \citet{uribe} to solve the stellar membership in open clusters, and \citet{SamuelLucas} and \citet{LucasPres08,LucasList,LucasPres11} to discover new star clusters in the UKIDSS GPS survey. Other applications in astronomy are by \citet{martinez} to estimate the power spectrum of the cosmic microwave background, and by \citet{yuan} for the calibration of a high resolution spectrometer.

\subsection{The algorithm}\label{subsec:algo}
The catalogue data is treated using a mixture model consisting of two-dimensional Gaussian densities to model the stellar clusters and of
homogeneous Poisson background to model the stars not belonging to the clusters.
\begin{equation}\label{mixModel}
P(X|\mu,\Sigma,\tau)=\prod_{i=1}^{N}\left [ \frac{\tau_0}{A}+\sum_{k=1}^{K}\tau_kp(x_i|k) \right ] \,,
\end{equation}
where $N$ is the number of sources within region $A$, $K$ the number of modeled clusters, $X$ the catalogue sources $\left \{ x_1,...,x_N \right \}$,
and for Gaussian clusters the multivariate normal Gaussian density is
\begin{equation}\label{Gaussians}
\begin{array}{l}
p(x_i|k)=\Phi_k(x_i|\mu_k,\Sigma_k)= \\\\
\frac{\displaystyle 1}{\displaystyle 2\pi\sqrt{\left | \Sigma_k \right |}}\textup{exp}\left [ -\frac{\displaystyle 1}{\displaystyle 2}(x_i-\mu_k)^T\Sigma_k^{-1}(x_i-\mu_k) \right ] \,.
\end{array}
\end{equation}

The model has three parameters: the mixing coefficients $\tau$ for the Gaussian clusters and the noise, and the means $\mu$ and covariances $\Sigma$ for the Gaussian clusters. Coefficient $\tau_0$ gives the proportion of stars belonging to the background and $\tau_i$ gives the proportion of stars belonging to the \textit{i}th cluster: $\sum_{k=0}^{K}\tau_k=1$.

After initializing the model parameters (see Sect. \ref{subsec:autom} step \ref{initCl}), the EM-algorithm works by repeating two alternating steps, the E-step and M-step. The E-step evaluates the \emph{responsibilities} i.e. the posterior probabilities of each point $x_i$ belonging to each group $k$ using the current parameter values.
\begin{equation}\label{Estep}
\begin{array}{l}
\Upsilon(z_{ik})=\frac{\displaystyle \frac{\tau_0}{A}}{\displaystyle \frac{\tau_0}{A}+\sum_{j=1}^{K}\tau_j\Phi(x_i|\mu_j,\Sigma_j)} \textup{ for }k=0 \\\\
\Upsilon(z_{ik})=\frac{\displaystyle \tau_k\Phi(x_i|\mu_k,\Sigma_k)}{\displaystyle \frac{\tau_0}{A}+\sum_{j=1}^{K}\tau_j\Phi(x_i|\mu_j,\Sigma_j)} \textup{ for }k>0
\end{array}
\end{equation}

The M-step re-estimates the parameters using the current responsibilities
\begin{equation}\label{Mstep}
\begin{array}{l}
\mu_k^{new}=\frac{\displaystyle 1}{\displaystyle N}\displaystyle\sum_{i=1}^N\Upsilon(z_{ik})x_i \textup{ for }k>0 \\\\
\Sigma_k^{new}=\frac{\displaystyle 1}{\displaystyle N}\displaystyle\sum_{i=1}^N\Upsilon(z_{ik})(x_i-\mu_k^{new})(x_i-\mu_k^{new})^T \textup{ for }k>0 \\\\
\tau_k^{new}=\frac{\displaystyle N_k}{\displaystyle N} \textup{ for }k\geq0
\end{array}
\end{equation}

where $N_k=\displaystyle \sum_{i=1}^N\Upsilon(z_{ik})$.

After each iteration round the \emph{log likelihood}, that indicates how well the current model parameters fit the data, is evaluated:
\begin{equation}\label{likehood}
l(X|\Theta)=\textup{ln}(P(X|\mu,\Sigma,\tau))=\sum_{i=1}^N\textup{ln}\left (\frac{\tau_0}{A}+\sum_{k=1}^K\tau_kp(x_i|k)\right ) \,.
\end{equation}
The E- and M-steps are repeated until convergence is reached for the log likelihood.\\\\
The above formulation in Eqs. \ref{mixModel}$-$\ref{likehood} of the mixture model and its estimation with the EM-algorithm is as in \citet{FraleyRaftery}, \citet{PRML}, and \citet{MCM}.

In our model we fix the number of Gaussian clusters $K$ to 1: we search smaller bins of data for one cluster at a time. We choose the diagonal covariance over the spherical and full covariance. The shape of a stellar cluster is usually not a circle. On the other hand the full covariance tends to suggest strong clusters in case of sparsely populated data bins or it might trap diffraction patterns of bright stars (see point \ref{spikeHalo} in Sect. \ref{appA1}) or other beam-like artefacts.

For the diagonal covariance Eq. (\ref{Gaussians}) changes to
\begin{equation}\label{diagGaussians}
\begin{array}{l}
p(x_i)=\Phi(x_i|\mu,\sigma)= \\\\
\frac{\displaystyle 1}{\displaystyle 2\pi\sqrt{\sigma_{xx}\sigma_{yy}}}\textup{exp}\left [ -\frac{\displaystyle 1}{\displaystyle 2}\left (\frac{\displaystyle (x_{ix}-\mu_{x})^2}{\displaystyle \sigma_{xx}} + \frac{\displaystyle (x_{iy}-\mu_{y})^2}{\displaystyle \sigma_{yy}}\right ) \right ] \,
\end{array}
\end{equation}
and for the covariance matrix
\begin{equation}\label{covMatrix}
\nonumber\Sigma=\begin{pmatrix} \sigma_{xx} & \sigma_{xy} \\ \sigma_{yx} & \sigma_{yy} \end{pmatrix}
\end{equation}
in Eq. (\ref{Mstep}) we have $\sigma_{xy}=\sigma_{yx}=0$ and
\begin{equation}\label{diagCov}
\begin{array}{l}
\sigma_{xx}^{new}=\frac{\displaystyle 1}{\displaystyle N}\displaystyle\sum_{i=1}^N\Upsilon(z_{i})(x_{ix}-\mu_{x}^{new})^2 \\\\
\sigma_{yy}^{new}=\frac{\displaystyle 1}{\displaystyle N}\displaystyle\sum_{i=1}^N\Upsilon(z_{i})(x_{iy}-\mu_{y}^{new})^2 \\
\end{array}
\end{equation}

\subsection{Automated search}\label{subsec:autom}

Clustering points in a two-dimensional space with the Gaussian mixture model is straightforward. However, the challenge in the UKIDSS GPS case is that searching for spatial densities among all catalogue data points without any other considerations locates clusters very poorly. Even if the data artefacts were filtered out it is likely that the clustering algorithm can locate only spatially compact star rich clusters if no additional data filtering takes place. This is because of the observed high spatial stellar density, strongly modulated by interstellar extinction, in the galactic plane. Sparse extended clusters do not sufficiently raise the spatial stellar number density to be caught by the algorithm. Therefore, suitable criteria must be applied to the data before clusters are searched for, i.e. the data must be pre-filtered.

Clustered star formation, be it low- or high-mass stars, takes place in dense molecular clouds. The dust in these clouds reflects the light from the newly born stars and this can be observed as localised surface brightness. If the stars are still embedded in the parental cloud or if they are obscured by foreground dust clouds the surface brightness is best observed in the \textit{K} band in which the interstellar extinction is the smallest of the UKIDSS filters. The presence of surface brightness is however not a proof of embedded clustered star formation. \textit{K} band surface brightness can also be due to e.g. formation of an embedded single star, planetary nebula or scattering of the interstellar radiation field from dust clouds (see e.g \citet{Lehtinen} and \citet{Juvela06,Juvela08}). \textit{K} band surface brightness can also be due to line emission from shocked molecular hydrogen resulting from interaction of molecular outflows from newly born low mass stars with the surrounding molecular cloud. Searching for surface brightness offers means to detect embedded star formation and stellar clusters.

The WSA catalogue data table \texttt{gpsSource} used in this study lists magnitudes of stars and galaxies detected in the survey but not explicit information on surface brightness. The catalogue parameter \texttt{mergedClass} is given for every object: -3 for a probable galaxy, -2 for a probable star, -1 for a star, 1 for a galaxy, and 0 for noise.

The star/galaxy classification is based on the object image profile used by the pipeline source detection algorithm (Irwin et al. in preparation). The objects with intensity profiles similar to the UKIDSS WFCAM point spread function are classified as stars, the rest as galaxies (or noise). Scrutiny of the data base and the survey images reveals that the detection algorithm tends to classify most of the objects within regions of variable surface brightness as galaxies. Thus the classification is more precisely star/non-stellar than star/galaxy.

The pipeline feature of classifying objects seen superposed on variable surface brightness as galaxies can be utilised in the search of stellar clusters either embedded in or near molecular/dust clouds. Besides the clusters also single embedded stars with associated nebulosities, either due to outflow activity or reflection, will produce "cluster" detections. Even though the galactic plane is in the centre of the zone of avoidance a large number of extragalactic sources are seen in the catalogue. These are also, quite rightly, classified as galaxies.

A fraction of the catalogue sources are due to data artefacts. These artefacts are discussed in detail in Irwin et al. (in preparation) and Appendix 2 of \citet{GPSdoc}. The artefacts cause highly varying extended surface brightness which causes the pipeline to classify most of the sources within the artefact as \texttt{mergedClass} $=+1$ sources. In addition sharp features in the artefacts produce non-existent \texttt{mergedClass} $=+1$ sources. The data artefacts must be filtered out from the data before the EM-algorithm is applied as otherwise too many false clusters due to artefacts will be located. 

The following artefacts have been addressed: Diffraction patterns of bright stars and diffraction spikes due to secondary mirror supports, bright stars at or near the border of the detector array, beams, 'bow-ties', cross-talk images and persistence images. The steps taken to allow for these artefacts are described in detail in online Appendix \ref{appA1}.

The classification of sources fainter than $17^\mathrm{m}$ in \textit{K} as star/non-stellar objects is highly unreliable. These sources were filtered out from the data.

The parameter \texttt{k\_1ppErrBits} contains the quality error information for each source detection of the \textit{K} filter. We accept sources with \texttt{k\_1ppErrBits} $<524288$. The value $131072 \leq$ \texttt{k\_1ppErrBits} $<524288$ for the \textit{K} band quality error bit flag refers to a GPS photometric calibration problem, that has been fixed in DR8 \citep{qualityErr}. This value of \texttt{k\_1ppErrBits} falls under the severity category of "Important Warning", but true sources can be found within such regions (e.g. clusters 107, 153, 154 and 155 in the list by \citet{LucasList}). We recognise that stars with \texttt{k\_1ppErrBits} $\geq 65536$ (the value $65536$ corresponds to "close to saturated") are considered to have unreliable photometry. In order not to loose true positives we apply the higher limit of \texttt{k\_1ppErrBits} $<524288$.

The catalogue data table \texttt{gpsSource} contains 125 attributes for each detected object. We tested the usefulness of other parameters in our clustering effort, but ultimately our method makes use only of the star/non-stellar classifier \texttt{mergedClass}.

%***********************************************************************
\begin{figure*}
\centering
\includegraphics[width=16cm]{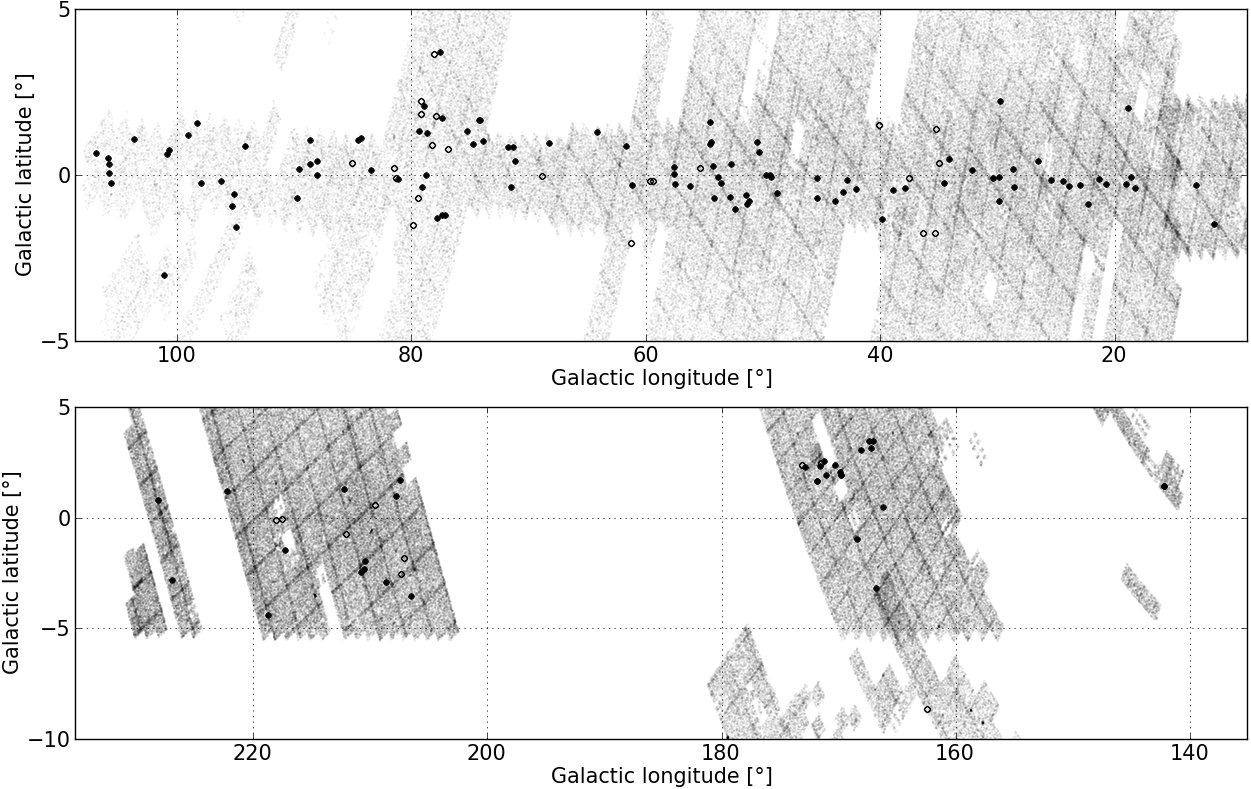}
\caption{Galactic distribution of the 137 cluster candidates (filled circles) in Table \ref{newClusters} and the 30 star formation location candidates (open circles) in Table \ref{newSFRs}.
The grey area marks the UKIDSS DR7 \textit{K} filter coverage. Only the northern edge of the Taurus-Auriga-Perseus star formation complex below the Galactic plane at l $\sim 170\degr$ is shown in the lower figure. No cluster candidates and only one star formation location candidate were found in this area.}\label{distrIm}% 
\end{figure*}
%***********************************************************************

UKIDSS DR7 contains 631\,117\,002 sources measured in the \textit{K} filter. Out of these 343\,737\,754 i.e. 54\% satisfy the criteria \textit{K} magnitude brighter than $17^\mathrm{m}$ and \texttt{k\_1ppErrBits} $<524288$. These sources are divided according to the \texttt{mergedClass} parameter so that a negligible fraction are probable galaxies or noise, 5\% probable stars, 74\% stars, and 19\% galaxies. We end up using for the detection algorithm sources with \textit{K} magnitude brighter than $17^\mathrm{m}$, \texttt{k\_1ppErrBits} $<524288$ and \texttt{mergedClass} $=+1$. These amount to 66\,149\,194 sources ($\sim10\%$ out of all sources in UKIDSS DR7). Besides for excluding objects with \textit{K} magnitude fainter than $17^\mathrm{m}$ and point \ref{magSteps} below the magnitudes listed in the UKIDSS catalogue are in no way used in the automated search.\\\\
The automated search proceeds in the following steps. Only the \textit{K} band data is used in the search.
   \begin{enumerate}
      \item The pre-filtered catalogue data is divided into smaller overlapping spatial bins of size 4\arcmin\ by 4\arcmin.
       Apart from bins at the dataset edges each bin overlaps one half of its neighbouring bins.
       4\arcmin\ by 4\arcmin\ was chosen as a suitable size for the bin based on experiments with the cluster candidates in the list by \citet{LucasList}.

      \item Remove false \texttt{mergedClass} $=+1$ classifications around bright stars and in the direction of the 8 diffraction spikes as explained in online Appendix \ref{appA1}.

      \item \label{magSteps} In order to track clusters with bright members the detection algorithm is run five times:
       once with all (filtered) input data and then using 80, 60, 40 and 20\% of these sources arranged in descending order of the \textit{K} magnitude.

      \item The spatial coordinates are rescaled to the interval [0,1] to make all bins equally important but still allowing them to have differing means and variances.
       This step is relevant only for bins at the dataset edges and which are smaller than 4\arcmin\ by 4\arcmin.

      \item\label{initCl} In order to initialise the model parameters the data bin is divided into 16 subgrids to find the area with the highest spatial density.
       The initial value of the cluster mean $\mu$ is the center point of the subgrid with the highest density.
       The covariance matrix of the data points assigned to the subgrid with the highest density give the initial values for the cluster covariance $\Sigma$.
       The weights $\tau$ have as initial values the same value: $\tau_0 = \tau_1 = 0.5$.

      \item Each data bin is represented by a mixture model of a background component and one Gaussian cluster component according to Sect. \ref{subsec:algo}.

      \item The EM-algorithm returns for each data bin a candidate cluster, i.e. an ellipse with the center point at the mean $\mu$ and
      half-axes determined by the covariance $\Sigma$.

      \item Remove false positives created by bright stars at or just outside an array edge as explained in online Appendix \ref{appA1}.

      \item Rearrange the candidates in descending order of the Bayesian information criterion (BIC, \citet{schwarz}).
      The BIC is used for rough comparison between competing models, and is defined as
        \begin{equation}
        \textup{BIC}=2l(X|\Theta)-d\textup{ln}(n),
        \end{equation}
        where $d$ denotes the number of degrees of freedom of the model,
        and $n$ is the number of data points.
        For this model with background noise and one Gaussian cluster with a diagonal mode covariance matrix, $d$ sums to 5:        
        \begin{itemize}
        \item Weights $\tau_0$ and $\tau_1$ with one constraint of normality: $\tau_0+\tau_1=1$.
        \item Means $\mu_x$ and $\mu_y$.
        \item Elements of the covariance matrix $\sigma_{xx}$ and $\sigma_{yy}$.
        \end{itemize}

      \item Merge cluster candidates closer than one arcmin to each other.

      \item Remove from the list the cluster candidates catalogued in \citet{BDBarbuy,BDSoares} (165 in UKIDSS DR7), \citet{MCM} (25 in DR7), \citet{FSR} (168 in DR7) and \citet{LucasList} (331 cluster candidates from DR4).

   \end{enumerate}

\subsection{Source screening}

Images of the cluster candidate areas with BIC $>$ 20 were retrieved from the database for visual inspection. Choosing 20 as the threshold value for the BIC gives 27599 cluster candidates which is a feasible number to inspect visually. Despite the effort to filter out data artefacts only $\sim2\%$ of these 27599 candidates are true cluster candidates, locations of embedded star formation, true galaxies or reflection nebulae. With the decreasing value of BIC the proportion of true candidates decreases strongly.

The cluster candidates were visually inspected and the obvious false clusters produced by data artefacts were excluded. The remaining candidates were screened by inspecting more thoroughly visually the grey scale \textit{J}, \textit{H}, \textit{K} and the false colour images (\textit{J} coded in blue, \textit{H} in green and \textit{K} in red) obtained from the WSA. The WSA images are automatically produced and the grey/colour scales are not necessarily optimised for resolving the high stellar densities in many of the clusters and the extended surface brightness in the locations of star formation. In such cases the \textit{J}, \textit{H} and \textit{K} fits files obtained from the WSA were used to produce grey scale and false colour images with better optimised intensity levels. Examples of such false colour images are shown in online Appendix \ref{appC}.

Visual inspection of many candidates revealed them to be galaxies or single stars with a reflection nebula. These we do not list. Such sources are expected since we search particularly for embedded clusters of non-stellar sources using the \texttt{mergedClass} $=+1$ criterion. Almost any object with surface brightness produces \texttt{mergedClass} $=+1$ classifications. Examples of false positive candidates are shown in online Appendix \ref{appA2}.

The selection criteria for a source to be accepted as a cluster or location of star formation candidate were conservative and doubtful objects were excluded. The selection was based solely on the optical appearance of the candidates and no attention was paid on the brightness of the stars in the direction of the candidates and no minimum number of stars in a possible cluster was required.

Finally SIMBAD data base was used to search for astronomical objects within 2\arcmin\ of the remaining candidates and SIMBAD literature links to these sources were inspected to exclude sources previously classified as stellar clusters or star forming locations. 

%__________________________________________________________________

\section{Results}\label{sec:results}

%***********************************************************************
\begin{figure*}
\centering
\includegraphics[width=16cm]{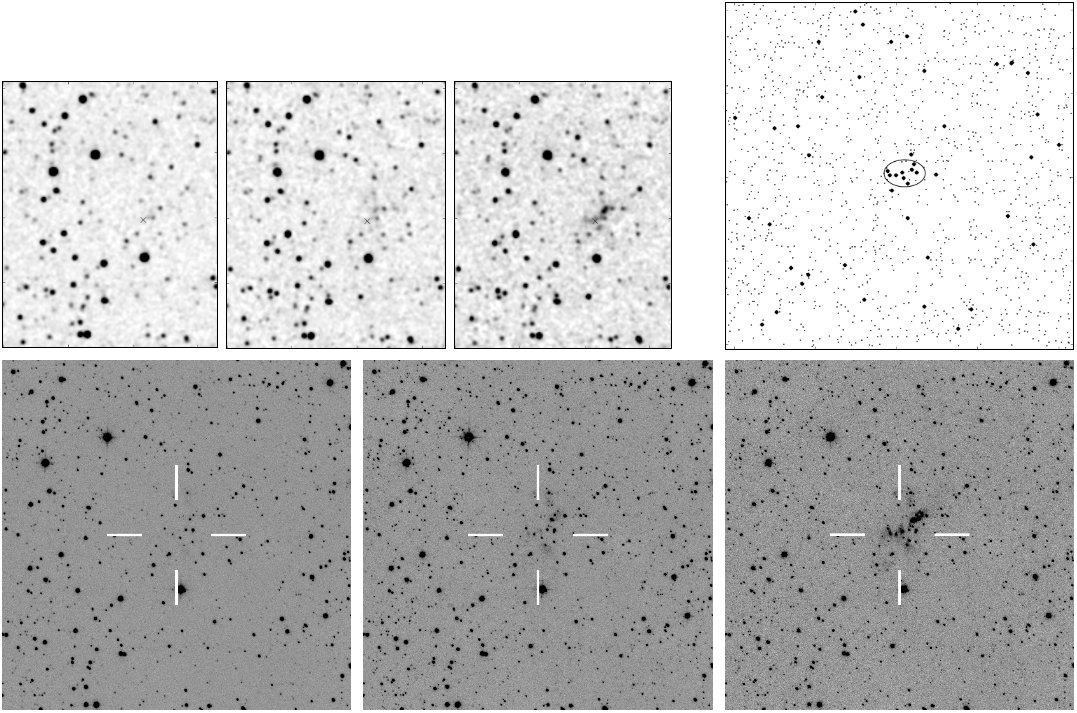}
\caption{Cluster candidate 80. In the lower panel 4\arcmin\ by 4\arcmin\ UKIDSS \textit{J}, \textit{H} and \textit{K} (from left to right) images of the cluster candidate area. Above the UKIDSS \textit{J} and \textit{H} images the same area in \textit{J}, \textit{H} and \textit{K} from 2MASS. Image orientation is North up and East left. Panel upper right: The UKIDSS \textit{K} band catalogue data. The non-stellar sources brighter than $17^\mathrm{m}$ in \textit{K} are plotted using large filled circles. The confidence ellipse provided by the EM algorithm is shown.}\label{clExample}%
\end{figure*}
%**********************************************************************

%***********************************************************************
\begin{figure*}
\centering
\includegraphics[width=16cm]{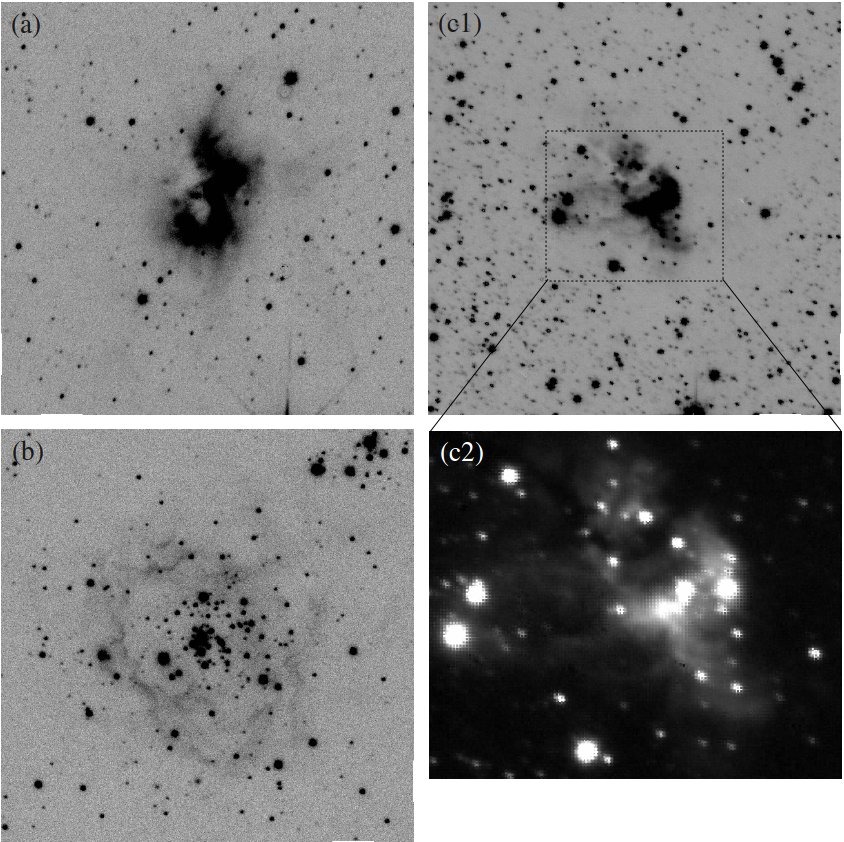}
\caption{Typical \textit{K} band images of cluster and star formation candidates. a) location of star formation candidate 15; b) cluster candidate 116; c1) and c2) cluster candidate 3. Image size is 4\arcmin\ by 4\arcmin\ and image orientation North up and East left.}\label{truePos}%
\end{figure*}
%***********************************************************************

The search located 137 cluster and 30 star formation location candidates which, to our knowledge, are previously unknown. The cluster candidates are listed in Table \ref{newClusters} and the candidate locations of star formation in Table \ref{newSFRs}. The columns list (1) a running number, (2) source identification, (3,4) Galactic coordinates, (4,5) J2000.0 equatorial coordinates, (6) description of selected SIMBAD sources within 2\arcmin\ of the direction of the candidate and (8) references to selected publications in Table \ref{Pubs}.

%********************************************************
\begin{figure*}
\centering
\includegraphics[width=16cm]{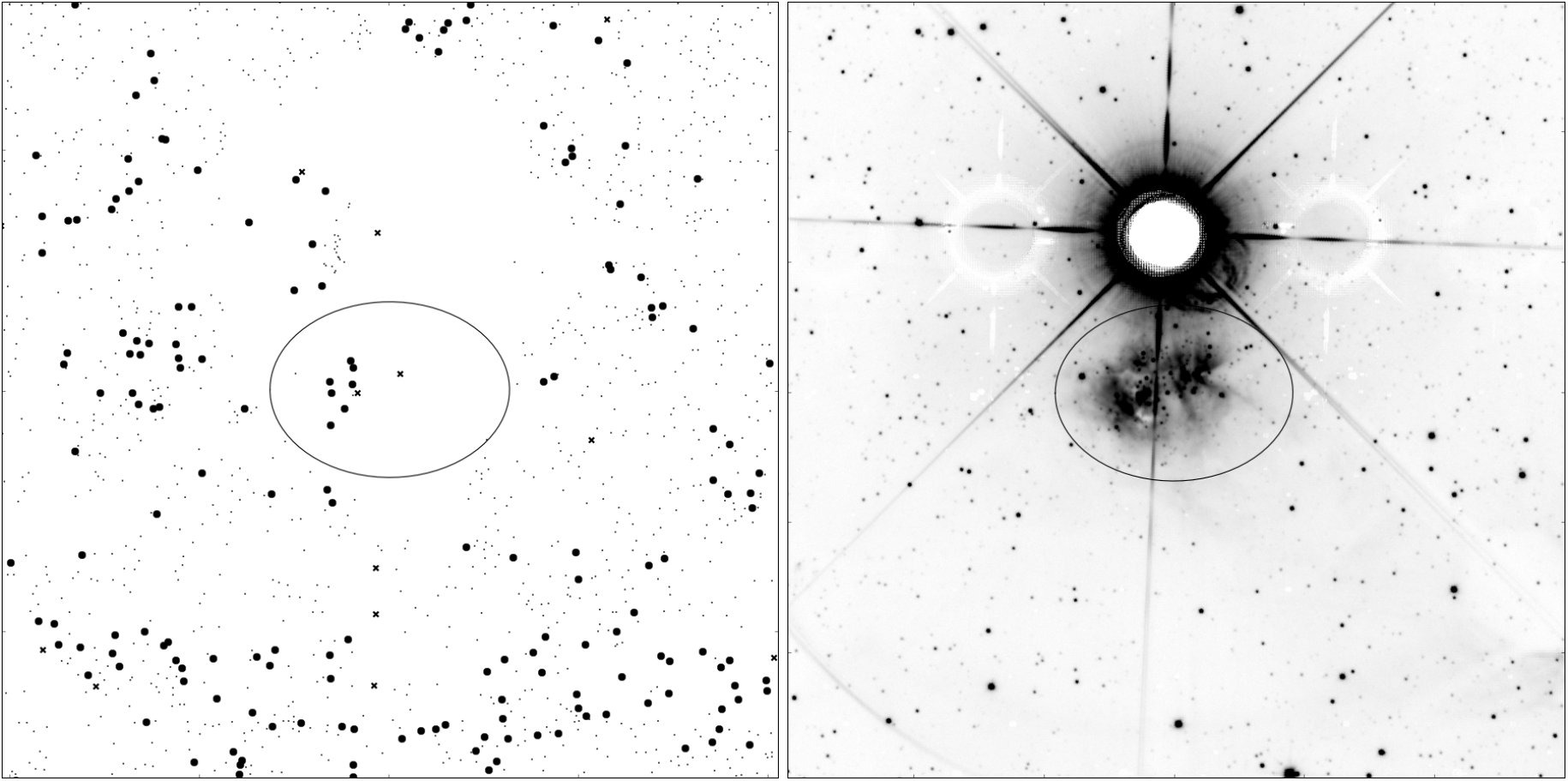}
\caption{The SH2-105 HII region ($l=75.834\degr,b=0.402\degr$) is located below a bright star. The catalogue data is shown in the panel on the left. The large filled circles are non-stellar sources brighter than $17^\mathrm{m}$ in \textit{K} and the crosses are sources that are listed in 2MASS but not in UKIDSS GPS. The 4\arcmin\ by 4\arcmin\ UKIDSS \textit{K} band image is shown in the panel on the right. The SH2-105 HII region is surrounded with an ellipse to indicate its position in the catalogue data plot.}\label{sh105}%
%K=0.74
\end{figure*}
%********************************************************

The distribution of the candidates is shown superposed on the observed GPS area in Fig. \ref{distrIm}. The cluster candidates are marked with filled circles and the star formation location candidates as open circles. The candidates are distributed quite symmetrically around the Galactic midplane. As expected most of the candidates lie within two degrees from the Galactic plane. The surface density of the candidates is higher in the direction of the inner Galaxy ($15\degr<l<107\degr$) than of the outer Galaxy ($141\degr<l<230\degr$). The area in the direction of the inner Galaxy includes 20 times more sources than of the outer Galaxy. Only the northern edge of the Taurus-Auriga-Perseus star formation complex below the plane scanned by the UKIDSS GPS is shown in Fig. \ref{distrIm}. No cluster candidates and only one star formation location candidate were found in this area.

4\arcmin\ by 4\arcmin\ images in \textit{JHK} bands of the new cluster candidate areas are available in electronic form\footnote{\url{http://www.helsinki.fi/~osolin/clusters}}. Most images show clear signs of reflected light in particular in the \textit{K} band thus indicating embedded clusters or sites of star formation.

Cluster candidate 80 is shown in Fig. \ref{clExample}. In the lower panel are the 4\arcmin\ by 4\arcmin\ UKIDSS \textit{JHK} images of the cluster candidate area. Reflected light from surrounding dust is visible in the \textit{K} image. Above the UKIDSS \textit{JH} images is the same area from 2MASS. A faint nebulosity can be spotted but no cluster. The cluster becomes visible as a spatial density when 60\% of the sources with \texttt{mergedClass} $=+1$ arranged in descending order of the \textit{K} magnitude (the large circles in the catalogue plot above the UKIDSS \textit{K} image) are fed to the algorithm. A millimetre radio source has been detected in this direction \citep{BGPS} but it has not been identified as a cluster.

Further \textit{K} band images of typical candidates are shown in Fig. \ref{truePos}. Location of star formation candidate 15 (Fig. \ref{truePos}a) has so far been identified only as an IRAS and a millimetre source. No stellar cluster is visible. Cluster candidate 116 (Fig. \ref{truePos}b) is visible as a galactic nebula in SDSS, but in the UKIDSS image there is a compact cluster. A second possibly associated cluster is seen NW of cluster candidate 116 in Fig. \ref{truePos}b. Cluster candidate 3 (Fig. \ref{truePos}c1) has around its location an IRAS source, an MSX source, an HII region, a submillimetre source, and a millimetre source. Expanded view of the cluster area using grey levels different from the WSA image is shown in Fig. \ref{truePos}c2. The cluster structure is better visible than in the image provided by WSA.

Further example cluster candidates including their colour-colour diagrams are shown in Appendix \ref{appB}.

%__________________________________________________________________
\section{Discussion}\label{sec:discussion}

Searching for spatial overdensities only in the number of stars in UKIDSS GPS is not fruitful. The number of stars in the Galactic plane is high and as a consequence sparse clusters do not increase the number of stars sufficiently to be detected. Also as strong extinction takes place in the Galactic plane the actually observed number of stars is highly modulated, and this modulation produces structures which trigger our model. One major culprit for the difficulty of an automated search for stellar clusters lies in the UKIDSS data base. From automated search point of view, the data base is plagued by strong clustered artefacts which overshadow real structures.

Straightforward clustering of all objects without filtering of the data fails, and therefore additional search criteria must be adopted. The requirement of associated surface brightness via the \texttt{mergedClass} $=+1$, i.e. non-stellar classification, chosen in this work directs the search to embedded stellar clusters. This criterion takes advantage of the UKIDSS catalogue feature of classifying stars superposed on variable background as non stellar objects thus producing clusters of such objects. Additionally, besides the clusters, the search targets also the locations of non-clustered star formation. Other criteria which were tested did not prove out successfully. The search was conducted in 16 square arcmin bins at a time which means that spatially extended clusters are not detected unless they are strongly centrally concentrated. We choose the diagonal covariance over the spherical and full covariance. Using the threshold value of 20 for the BIC gives 27599 candidates out of which $\sim2\%$ are regarded true positive candidates through visual inspection of the candidate images. Out of these 167 ($\sim30\%$) candidates are stellar clusters or sites of non-clustered star formation not previously verified as such. The EM method, as applied in this work, performs well with the \citet{BDBarbuy,BDSoares} and \citet{LucasList} catalogue objects that are compact enough to fit the 4\arcmin\ by 4\arcmin\ window, but very poorly with the catalogue of \citet{FSR} that was compiled by searching for statistical over-densities only.
Among the cluster candidates in the list by \citet{LucasList} found also by our system are both nebulae and clusters with some or no nebulosity. Cluster candidates in the list by \citet{LucasList} not found by our system were not found either because they did not represent themselves as clusters of non-stellar sources, or the BIC value given by our system fell under our cut-off value of 20.

Surface brightness due to embedded stellar clusters or star formation is only one indication of presence of such objects. Young, embedded stars are usually associated with infrared objects (e.g. IRAS or Spitzer), masers (e.g. H2O, SiO, methanol) and extended or point like (sub)mm sources. Early type stars are associated with HII regions. Numerous surveys for these star formation indicators have been conducted in the direction of e.g. colour selected IRAS sources. None of these indicators was used in the EM search. It is therefore of interest to study if one or more of these indicators have been detected in the direction of the new clusters and embedded star formation locations.

SIMBAD was used to search for sources within 2\arcmin\ from the cluster or embedded star formation candidates with the following results (the number of sources are given in parenthesis): IRAS point source (100), MSX source (38), (sub)millimetre source (60), maser (24), outflow candidate (4) and HII region (39). 31 candidates are seen in the direction of a Spitzer infrared dark cloud (IRDC). Cirrus-like IRAS point souces (IRAS detection only at 100 microns) were excluded. If the 100 micron IRAS flux was of low quality or an upper limit only a good quality flux rising from 12 microns to 60 microns was required. All the IRAS point sources listed as associated sources in Table \ref{newClusters} have IRAS fluxes rising from 12 microns to 100 microns, i.e. typical for embedded sources in star forming clouds. Mostly more than one of these indicators were seen in the direction of the candidates. 32 cluster candidates and 7 embedded star formation candidates were not associated with any object in the SIMBAD data base. The number of indicators seen in the direction of the candidates gives confidence that most of the new clusters or embedded star formation locations are real entities and not produced by chance nor are due to catalogue artefacts.
 
Although the EM-algorithm returns the half-axes of the ellipse covering the cluster candidate area, this method of clustering non-stellar sources produced by surface brightness is not adequate for deriving estimates for the cluster radii and the number of members. The radii could be estimated through visual examination of the images, but accurate estimates are outside the scope of this study.

The UKIDSS GPS covers the direction of the Galactic anti-centre. This region of sky is well visible from the northern hemisphere and has been intensively studied optically, in IR and in radio domain. The optical extinction in this general direction is also by far not so severe as the general direction of the Galactic centre. One would therefore not expect to find many previously undetected large and star rich clusters. Contrary to expectations a number of such clusters were detected (candidates 106, 107, 109, 110, 103, 114, 116 and 126). Of these the clusters pairs 106-107, 115-116 and 122-123 are seen in the same 4\arcmin\ by 4\arcmin\ image. Clusters 107 and 109 have many different SIMBAD indicators of star formation seen in their direction, 114 an IRAS point source and an HII region, 116 and 126 an IRAS point source and reflection nebulosity. Cluster 115 has no associated indicator. The small apparent size indicates that these clusters lie far in the outer Galactic plane. IRAS point sources in the direction of cluster candidates 107-110, 114-120, 124-128, 130, 132, 134, 136 and 137 and location of star formation candidates 25, 26, 28 and 30 have been included in the CO survey of \citet{WB89}, who have extensively studied the star formation in the outer Galaxy during the past two decades. Further detailed study of these sources will shed light on the star formation history of the outer Galaxy.

Further insight to the cluster candidates can be obtained by investigating their $(H-K, J-H)$ colour-colour diagrams. Images and colour-colour diagrams of selected cluster candidates are shown in online Appendix \ref{appB}, Figs. \ref{cc43} to \ref{cc114}. The colour-colour diagram is a useful tool to investigate the cluster properties and membership of individual stars if the photometric data is accurate. The background surface brightness and crowded stellar fields in the direction of many of the new cluster candidates makes accurate photometry, especially for the faint stars, difficult. UKIDSS sources brighter than 17 magnitudes in \textit{K} and classified as non-stellar were used to find the cluster candidates. In the following "cluster indicator" refers to UKIDSS sources (both stellar and non-stellar) in the cluster direction brighter than $17^\mathrm{m}$ in \textit{K}. Of the example clusters in online Appendix B for cluster candidates 114 and 116 a large fraction of the cluster indicators lie within the reddening band. For the other candidates either a large (cluster candidates 20 and 110) or a major fraction of the cluster indicators lie to the right (cluster candidates 9 and 43) of the reddening band. The position of objects right of the reddening band could be explained by infrared excess caused by circumstellar dust but such a high number of these stars in the clusters is not expected. According to the colour-colour diagrams the foreground extinctions towards the cluster indicators are up from 10 magnitudes (early spectral type assumed) up to 30 or more magnitudes. This is a reasonable value. If the extinctions were significantly lower these clusters would have already been detected in the optical. Dedicated NIR imaging of these clusters is needed to obtain accurate cluster indicator photometry.

\subsection{Notes on individual candidates}\label{subsec:individ}
{\bf Cluster candidate 5} is seen at the edge of a dense dust cloud.\\
{\bf Cluster candidate 18}: the nebulous object seen to NW of the cluster position of this candidate is associated with a methanol maser that has been listed in three studies: \citet{Caswell}, \citet{BK2004} and \citet{SKH2002}. The maser from the Caswell list is used as an example on pp.17$-$20 in a presentation by \citet{LucasPres08}.\\
{\bf Cluster candidate 29}: the nebulous structure southwest is probably associated.\\
{\bf Cluster candidate 71}: \object{[BDB2003] G077.46+01.76} is 2.8\arcmin\ away from this candidate.\\
{\bf Cluster candidate 121}: 3.1\arcmin\ southwest of this candidate is a concentration of stars.\\
{\bf Cluster candidate 131}: cluster candidate nro 15 in the list by \citet{LucasList} is 5.1\arcmin\ away from this candidate.\\
{\bf Location of star formation candidate 3}: the HII region \object{W 48C} is 1.8\arcmin\ away from this candidate.\\
{\bf Location of star formation candidate 6}: the \object{SH2-75} HII region and an IRAS source classified as a cluster by \citet{Faustini} is 2.5\arcmin\ NE of this candidate. The HII region has a diameter of 10\arcmin\ making this candidate part of SH2-75. No stellar cluster is visible in the image, but instead an object that could be e.g. an outflow cone.\\
{\bf Location of star formation candidates 12} and {\bf 15} and {\bf cluster candidate 128} are possible molecular hydrogen objects based on UKIDSS images. \\
{\bf Location of star formation candidate 25}: two other nebulosities are seen nearby this candidate. \\
{\bf Location of star formation candidate 29}: northeast of this candidate are a second nebulous source 1.7\arcmin\ away and cluster candidate \object{[IBP2002] CC09} in \citet{IBP2002} 3.2\arcmin\ away.\\
{\bf Location of star formation candidate 30}: the \object{SH 2-287 A} HII region. \\

With the exception of cluster candidates 5, 17, 23, 56, 63, 66, 69, 72, 75, 76, 81, 82, 83, 86, 93, 97, 98, 99, 102, 103, 105, 111, 113, 115, 119, 121, 122, 125, 129, 130, 131, 133, 135 and 136 and location of star formation candidates 2, 6, 11, 12, 17, 19, 25 and 29 the rest of the candidates are associated at least with an IRAS point source, most also with other indicators of star formation. Many of our candidates are included in various studies:
\begin{itemize}
\item Classified as star forming regions (SFRs) based on sub-mm continuum imaging of IRAS sources selected from radio ultracompact HII region surveys \citep{THW}. (Cluster candidates 3, 6, 14, 15, 24 and 29 and location of star formation candidates 1 and 5).
\item Suspected sites of massive star formation based on millimetre continuum emission survey toward regions previously identified as harbouring a methanol maser and/or a radio ultracompact HII region \citep{HBM}. (Cluster candidates 1, 3 and 18 and location of star formation candidates 1 and 5).
\item Included in a millimetre continuum and CS spectral line study of massive star forming regions in very early stages of evolution, most of them prior to building up an ultracompact HII region \citep{BSM}. (Cluster candidates 4, 12, 19, 20 and 27 and location of star formation candidate 8).
\item Included in a study of 850 $\mu$m and 450 $\mu$m continuum emission seen towards a sample of high-mass protostellar objects (HMPOs) \citep{WFS}. (Cluster candidates 4, 12, 19, 20 and 26 and location of star formation candidate 8).
\item Included in the APEX submillimetre survey that searches for massive pre- and proto-stellar clumps in the Galaxy in order to shed light on the early stages of star formation \citep{SMC2009}. (Cluster candidates 4 and 6).
\item Identified as Extended Green Objects in a mid-IR survey by their extended 4.5 $\mu$m emission that may be an indicator of outflows specifically from massive protostars \citep{EGO}. (Cluster candidates 19, 48 and 54 and location of star formation candidate 1).
\item Included in a submillimetre survey whose primary goal is to identify and characterise HMPOs \citep{CAB2008} (Location of star formation candidates 8 and 9). The region studied is near the open cluster NGC 6823 to which candidate 8 has a distance of 4.3\arcmin\ and candidate 9 15\arcmin.
\item Bubble candidates from GLIMPSE \citep{CPA2006}. (Cluster candidates 8, 27, 28, 31, 33 and 53).
\item Identified as zone of avoidance galaxies \object{ZOAG G166.81-03.20}, \object{ZOAG G167.06+03.46}, \object{ZOAG G167.42+03.45} \citep{zoagIV} and \object{ZOAG G228.10+00.80} \citep{zoagII}. (Cluster candidates 109, 110, 112 and 137).
\end{itemize}

In general radio surveys find circumstellar dust envelopes and disks, and cold cores of molecular clouds. In areas where a radio telescope sees only a point source or signs of e.g. an ultracompact HII region, the UKIDSS images show structures of surface brightness and single stars thus verifying the results of the millimetre/submillimetre radio surveys of suspected star forming regions.

Zone of avoidance galaxies (ZOAG) have been identified in the direction of four of the new cluster candidates (109, 110, 112 and 137). False colour images produced using the WSA fits files in online Figs. \ref{appC109}, \ref{appC110}, \ref{appC112} and \ref{appC137} (cluster candidate 110 is presented also in Fig. \ref{cc110} in Appendix \ref{appB}) show that instead of being extragalactic sources they are Galactic clusters. A cluster of individual stars are seen in all figures. This would not be the case if the objects were extragalactic.

The discovery of the UKIDSS stellar cluster in the direction of \object{SH2-105} was serendipitious. It is not surprising to detect a stellar cluster in a HII region. On the contrary, it would be surprising not to find one. The cluster was discovered while inspecting the effect of bright stars to the GPS catalogue. The UKIDSS catalogue data is plotted in Fig. \ref{sh105}. The large filled circles are non-stellar sources brighter than $17^\mathrm{m}$ in \textit{K} and the crosses are sources that are listed in 2MASS but not in UKIDSS GPS. The UKIDSS \textit{K} band image is shown in the right panel. Even though an obvious cluster is seen in the \textit{K} band image practically no stars are listed in the catalogue. The presence of the bright star prevents automated star detection and thus produces a void into the catalogue. The small "cluster" NE of the bright star in the UKIDSS catalogue data is produced mainly by the diffraction rings around the star. The serendipitous detection of the SH2-105 cluster indicates that GPS images may hold many objects, clusters or other, which can not be found using the GPS catalogue because the data is either missing or corrupted.

\subsection{Summary}

While the mixture model used here is an effective method to automatically locate clusters in a large amount of data, the ratio of true positive to false positive candidates given by our system, even though the input data was heavily filtered, is still poor. The processing of the data before it is given to the algorithm is at this stage quite limited: we start by choosing the non-stellar sources and proceed with removing potential false positives before and after the algorithm gives a list of candidates. Also judging from all the false positive examples presented here the \texttt{mergedClass} stellar/non-stellar classifier is often unreliable. This is to be expected within nebulous regions and in the vicinity of very bright stars, where the surface brightness also has a steep gradient. Also such classifiers are unreliable for low signal to noise ratio detections and marginally resolved stellar pairs are often mis-classified as a single non-stellar source.

%______________________________________________________________

\section{Conclusions}\label{sec:conclusions}

We have used Gaussian mixture modelling, optimised with the Expectation Maximization algorithm to locate embedded stellar clusters and locations of star formation from the UKIDSS Galactic Plane Survey data release 7. Taking advantage of a feature of the UKIDSS stellar classification method which tends to classify stars superposed on variable surface brightness as non-stellar objects we have targeted clusters associated with enhanced sky surface brightness, i.e. mainly embedded clusters. Approximately 10\% (66 million objects) of the UKIDSS GPS DR7 objects in \textit{K} band brighter than $17^\mathrm{m}$ are classified non-stellar. However, the UKIDSS catalogue artefacts due to e.g. bright stars mimic true high surface brightness areas and produce clusters of objects classified as non-stellar by the UKIDSS pipeline. Without a proper filtering image artefacts strongly overshadow true clusters in an automated search. Despite heavy filtering only a few percent of the cluster candidates produced by the automated search turn out not to be data artefacts or false positives. Besides clusters also a number of candidates for locations of star formation were found. The real clusters and locations of star formation had to be visually selected from the list suggested by the automated search.

After discarding the already known clusters and protostars 137 previously unknown stellar clusters and 30 locations of star formation were found. An IRAS point source is seen in the direction of most of the new clusters and locations of star formation. An IRAS source is considered to be associated with a candidate only if it is close enough to the candidate, and the flux density increases towards 100 microns. Besides the IRAS point sources other indications of a still ongoing star formation (e.g. (sub)mm, MSX and maser sources) are detected in the direction or near a large part of the detected clusters. As expected most of the detected clusters or star formation locations are tightly concentrated on the Galactic plane. Relatively few clusters were detected in the direction of the northern Galactic plane. This possibly indicates that most of the northern clusters have already been discovered as this part of the plane has been more thoroughly investigated than the southern plane. However, some of the new northern clusters in the direction of the Galactic anticentre are massive and deserve to be investigated in more detail.

We will continue our search with the future UKIDSS releases. The part of the galactic plane which is not visible at Mauna Kea is being surveyed in the NIR by the VISTA telescope at Paranal observatory. Search for southern clusters down to the same limiting magnitude as the UKIDSS data will thus be possible in near future.

\begin{acknowledgements}
This work was funded by the Finnish Ministry of Education under the project "Utilizing Finland's membership in the European Southern Observatory". This work was supported by the Academy of Finland under grants 118653 (ALGODAN) and 132291, and by the Finnish Funding Agency for Technology and Innovation (TEKES) under the project MIFSAS. This work uses data products from the Two Micron All Sky Survey, and the United Kingdom Infrared Telescope Infrared Deep Sky Survey. This research uses the SIMBAD astronomical database service operated at CCDS, Strasbourg. We thank the referee Philip Lucas for useful comments and suggestions.
\end{acknowledgements}

\bibliographystyle{aa} % style aa.bst
\bibliography{arXiv-aa18531-11}

\listofobjects

\tiny
\begin{longtable}{l*{8}{c}}
\caption{\label{newClusters}List of cluster candidates.}\\
\hline\hline
\# & ID & $l$ & $b$ & $\alpha$\ (J2000) & $\delta$\ (J2000) & Associated sources\tablefootmark{S} & References \\
 & & [$\degr$] & [$\degr$] & [$h$\ $m$\ $s$] & [$\degr$\ ' \ ``]  &  &  \\
\hline
\endfirsthead
\caption{continued.}\\
\hline\hline
\# & ID & $l$ & $b$ & $\alpha$\ (J2000) & $\delta$\ (J2000) & Associated sources\tablefootmark{S} & References \\
 & & [$\degr$] & [$\degr$] & [$h$\ $m$\ $s$] & [$\degr$\ ' \ ``]  &  &  \\
\hline
\endhead
\hline
\endfoot
            1 & G011.495$-$1.483 & 11.495 & $-$1.483 & 18 16 21 & $-$19 41 31 & IRAS,MSX,HII,smm,mm,Mas & 3,6,11,12,17 \\
            2 & G013.076$-$0.309 & 13.076 & $-$0.309 & 18 15 11 & $-$17 44 35 & IRAS,mm,IRDC & 18 \\
            3 & G018.303$-$0.392 & 18.303 & $-$0.392 & 18 25 43 & $-$13 10 23 & IRAS,MSX,HII,smm,mm & 2,3 \\
            4 & G018.655$-$0.059 & 18.655 & $-$0.059 & 18 25 11 & $-$12 42 22 & IRAS,MSX,smm,mm,IRDC & 4,5,7,18 \\
            5 & G018.850$+$2.023 & 18.850 & $+$2.023 & 18 18 02 & $-$11 33 18 & \ldots & \ldots \\
            6 & G019.073$-$0.286 & 19.073 & $-$0.286 & 18 26 48 & $-$12 26 31 & IRAS,MSX,HII,smm,mm,IRDC & 2,7,18 \\
            7 & G020.711$-$0.291 & 20.711 & $-$0.291 & 18 29 56 & $-$10 59 38 & IRAS,mm & \ldots \\
            8 & G021.343$-$0.137 & 21.343 & $-$0.137 & 18 30 34 & $-$10 21 47 & mm,IRDC,bub & 13,18 \\
            9 & G022.257$-$0.880 & 22.257 & $-$0.880 & 18 34 57 & $-$09 53 42 & IRAS & \ldots \\
            10 & G022.952$-$0.316 & 22.952 & $-$0.316 & 18 34 13 & $-$09 01 05 & IRAS,HII,mm,IRDC & 18 \\
            11 & G023.882$-$0.353 & 23.882 & $-$0.353 & 18 36 05 & $-$08 12 36 & IRAS,mm,IRDC & 18 \\
            12 & G024.397$-$0.188 & 24.397 & $-$0.188 & 18 36 27 & $-$07 40 34 & IRAS,smm,mm,IRDC & 4,5,18 \\
            13 & G025.462$-$0.159 & 25.462 & $-$0.159 & 18 38 19 & $-$06 43 01 & mm & \ldots \\
            14 & G026.544$+$0.414 & 26.544 & $+$0.414 & 18 38 16 & $-$05 29 35 & IRAS,MSX,HII,smm,mm,2MASX,IRDC & 2,18 \\
            15 & G028.592$-$0.365 & 28.592 & $-$0.365 & 18 44 49 & $-$04 01 41 & IRAS,HII,smm,mm & 2 \\
            16 & G028.693$+$0.177 & 28.693 & $+$0.177 & 18 43 04 & $-$03 41 28 & MSX,HII,mm,DNe & 8 \\
            17 & G029.815$+$2.224 & 29.815 & $+$2.224 & 18 37 50 & $-$01 45 25 & IRAS,MSX & \ldots \\
            18 & G029.858$-$0.060 & 29.858 & $-$0.060 & 18 46 02 & $-$02 45 47 & MSX,smm,mm,Mas,IRDC & 3,18 \\
            19 & G029.888$-$0.779 & 29.888 & $-$0.779 & 18 48 40 & $-$03 03 50 & IRAS,smm,mm,of?,IRDC & 4,5,9,18 \\
            20 & G030.385$-$0.107 & 30.385 & $-$0.107 & 18 47 10 & $-$02 18 58 & IRAS,MSX,HII,smm,mm,IRDC & 4,5,18 \\
            21 & G032.152$+$0.131 & 32.152 & $+$0.131 & 18 49 33 & $-$00 38 02 & IRAS,MSX,HII,mm,Mas,IRDC & 18 \\
            22 & G034.132$+$0.472 & 34.132 & $+$0.472 & 18 51 57 & $+$01 16 59 & IRAS,MSX,HII,mm & \ldots \\
            23 & G034.583$-$0.238 & 34.583 & $-$0.238 & 18 55 18 & $+$01 21 40 & \ldots & \ldots \\
            24 & G037.876$-$0.400 & 37.876 & $-$0.400 & 19 01 54 & $+$04 12 58 & IRAS,MSX,HII,smm,mm,Mas,IRDC & 2,18 \\
            25 & G038.937$-$0.459 & 38.937 & $-$0.459 & 19 04 04 & $+$05 07 55 & MSX,mm,DNe & 8 \\
            26 & G039.903$-$1.351 & 39.903 & $-$1.351 & 19 09 02 & $+$05 34 48 & HII,Mas & \ldots \\
            27 & G042.111$-$0.447 & 42.111 & $-$0.447 & 19 09 54 & $+$07 57 22 & IRAS,HII,smm,mm,bub & 4,5,13 \\
            28 & G042.834$-$0.151 & 42.834 & $-$0.151 & 19 10 11 & $+$08 44 02 & IRAS,smm,IRDC,bub & 13,18 \\
            29 & G043.186$-$0.525 & 43.186 & $-$0.525 & 19 12 11 & $+$08 52 23 & MSX,HII,smm,mm,Mas,IRDC & 1,2,6,18 \\
            30 & G043.889$-$0.784 & 43.889 & $-$0.784 & 19 14 26 & $+$09 22 34 & IRAS,MSX,HII,smm,Mas,IRDC & 1,6,18 \\
            31 & G045.397$-$0.709 & 45.397 & $-$0.709 & 19 17 01 & $+$10 44 42 & HII,IRDC,bub & 13,18 \\
            32 & G045.417$-$0.105 & 45.417 & $-$0.105 & 19 14 53 & $+$11 02 38 & IRAS,mm,IRDC & 18 \\
            33 & G048.845$-$0.544 & 48.845 & $-$0.544 & 19 23 03 & $+$13 52 05 & IRDC,bub & 13,18 \\
            34 & G049.288$-$0.056 & 49.288 & $-$0.056 & 19 22 08 & $+$14 29 17 & IRAS,mm,2MASX,Mas,IRDC & 17,18 \\
            35 & G049.430$-$0.011 & 49.430 & $-$0.011 & 19 22 15 & $+$14 38 06 & IRAS,MSX,mm,IRDC & 18 \\
            36 & G049.721$-$0.016 & 49.721 & $-$0.016 & 19 22 50 & $+$14 53 20 & IRAS,mm,IRDC & 18 \\
            37 & G050.317$+$0.675 & 50.317 & $+$0.675 & 19 21 28 & $+$15 44 24 & IRAS,MSX,HII,Mas & 1,6 \\
            38 & G050.490$+$0.994 & 50.490 & $+$0.994 & 19 20 38 & $+$16 02 35 & IRAS & \ldots \\
            39 & G051.210$-$0.799 & 51.210 & $-$0.799 & 19 28 37 & $+$15 49 37 & IRAS & \ldots \\
            40 & G051.401$-$0.890 & 51.401 & $-$0.890 & 19 29 20 & $+$15 57 07 & IRAS & \ldots \\
            41 & G051.426$-$0.615 & 51.426 & $-$0.615 & 19 28 23 & $+$16 06 18 & IRAS & \ldots \\
            42 & G052.367$-$1.044 & 52.367 & $-$1.044 & 19 31 50 & $+$16 43 30 & IRAS & \ldots \\
            43 & G052.753$+$0.335 & 52.753 & $+$0.335 & 19 27 32 & $+$17 43 26 & IRAS,MSX,HII,mm,IRDC & 18 \\
            44 & G052.847$-$0.664 & 52.847 & $-$0.664 & 19 31 24 & $+$17 19 41 & IRAS & \ldots \\
            45 & G053.594$-$0.249 & 53.594 & $-$0.249 & 19 31 23 & $+$18 10 59 & mm & \ldots \\
            46 & G053.819$-$0.059 & 53.819 & $-$0.059 & 19 31 08 & $+$18 28 19 & IRAS,mm,IRDC & 18 \\
            47 & G054.192$-$0.691 & 54.192 & $-$0.691 & 19 34 14 & $+$18 29 35 & IRAS & \ldots \\
            48 & G054.236$+$0.257 & 54.236 & $+$0.257 & 19 30 49 & $+$18 59 20 & IRAS & \ldots \\
            49 & G054.426$+$0.991 & 54.426 & $+$0.991 & 19 28 28 & $+$19 30 29 & of?,IRDC & 9,18 \\
            50 & G054.493$+$1.579 & 54.493 & $+$1.579 & 19 26 25 & $+$19 50 49 & IRAS & \ldots \\
            51 & G054.522$+$0.919 & 54.522 & $+$0.919 & 19 28 56 & $+$19 33 29 & IRAS,IRDC & 18 \\
            52 & G056.239$-$0.342 & 56.239 & $-$0.342 & 19 37 10 & $+$20 27 04 & IRAS & \ldots \\
            53 & G057.546$-$0.273 & 57.546 & $-$0.273 & 19 39 40 & $+$21 37 26 & IRAS,MSX,HII,mm,IRDC,bub & 13,18 \\
            54 & G057.573$+$0.221 & 57.573 & $+$0.221 & 19 37 52 & $+$21 53 24 & IRAS,IRDC & 18 \\
            55 & G057.608$+$0.024 & 57.608 & $+$0.024 & 19 38 41 & $+$21 49 26 & IRAS,mm,of? & 9 \\
            56 & G061.193$-$0.299 & 61.193 & $-$0.299 & 19 47 40 & $+$24 46 23 & \ldots & \ldots \\
            57 & G061.720$+$0.863 & 61.720 & $+$0.863 & 19 44 24 & $+$25 48 40 & IRAS,MSX,HII,2MASX & \ldots \\
            58 & G064.152$+$1.283 & 64.152 & $+$1.283 & 19 48 15 & $+$28 07 30 & IRAS,MSX,HII & \ldots \\
            59 & G068.239$+$0.960 & 68.239 & $+$0.960 & 19 59 13 & $+$31 27 47 & IRAS,MSX,HII & \ldots \\
            60 & G071.151$+$0.399 & 71.151 & $+$0.399 & 20 08 50 & $+$33 37 34 & IRAS,MSX,HII,smm,mm,2MASX & \ldots \\
            61 & G071.312$+$0.827 & 71.312 & $+$0.827 & 20 07 32 & $+$33 59 35 & IRAS,Mas & 11,12,17 \\
            62 & G071.523$-$0.386 & 71.523 & $-$0.386 & 20 12 58 & $+$33 30 29 & IRAS,MSX,mm,Mas & 6,11,17 \\
            63 & G071.804$+$0.846 & 71.804 & $+$0.846 & 20 08 44 & $+$34 25 01 & \ldots & \ldots \\
            64 & G073.878$+$1.026 & 73.878 & $+$1.026 & 20 13 34 & $+$36 15 04 & IRAS,MSX,HII,2MASX & \ldots \\
            65 & G074.159$+$1.645 & 74.159 & $+$1.645 & 20 11 46 & $+$36 49 34 & IRAS,smm,Mas & 17 \\
            66 & G074.213$+$1.650 & 74.213 & $+$1.650 & 20 11 54 & $+$36 52 26 & \ldots & \ldots \\
            67 & G074.753$+$0.913 & 74.753 & $+$0.913 & 20 16 27 & $+$36 54 58 & IRAS,MSX,HII,2MASX & \ldots \\
            68 & G075.295$+$1.324 & 75.295 & $+$1.324 & 20 16 16 & $+$37 35 42 & IRAS,HII,Mas & 17 \\
            69 & G077.127$-$1.228 & 77.127 & $-$1.228 & 20 32 07 & $+$37 37 30 & \ldots & \ldots \\
            70 & G077.405$-$1.213 & 77.405 & $-$1.213 & 20 32 54 & $+$37 51 29 & IRAS,mm & \ldots \\
            71 & G077.437$+$1.720 & 77.437 & $+$1.720 & 20 20 45 & $+$39 35 18 & MSX,HII & \ldots \\
            72 & G077.568$+$3.693 & 77.568 & $+$3.693 & 20 12 33 & $+$40 47 49 & \ldots & \ldots \\
            73 & G077.821$-$1.310 & 77.821 & $-$1.310 & 20 34 33 & $+$38 08 02 & IRAS,mm & \ldots \\
            74 & G078.703$+$1.243 & 78.703 & $+$1.243 & 20 26 35 & $+$40 21 04 & mm & \ldots \\
            75 & G078.734$-$0.021 & 78.734 & $-$0.021 & 20 32 01 & $+$39 38 05 & \ldots & \ldots \\
            76 & G078.920$+$2.058 & 78.920 & $+$2.058 & 20 23 44 & $+$40 59 53 & \ldots & \ldots \\
            77 & G079.134$-$0.368 & 79.134 & $-$0.368 & 20 34 42 & $+$39 45 00 & mm & \ldots \\
            78 & G079.378$+$1.324 & 79.378 & $+$1.324 & 20 28 18 & $+$40 56 49 & mm & \ldots \\
            79 & G081.123$-$0.133 & 81.123 & $-$0.133 & 20 40 03 & $+$41 28 37 & MSX,mm & \ldots \\
            80 & G083.464$+$0.155 & 83.464 & $+$0.155 & 20 46 41 & $+$43 29 42 & mm & \ldots \\
            81 & G084.286$+$1.101 & 84.286 & $+$1.101 & 20 45 25 & $+$44 43 37 & \ldots & \ldots \\
            82 & G084.528$+$1.053 & 84.528 & $+$1.053 & 20 46 29 & $+$44 53 10 & \ldots & \ldots \\
            83 & G088.056$-$0.017 & 88.056 & $-$0.017 & 21 04 17 & $+$46 53 13 & \ldots & 16 \\
            84 & G088.099$+$0.418 & 88.099 & $+$0.418 & 21 02 33 & $+$47 12 29 & mm,Mas & 17 \\
            85 & G088.660$+$1.032 & 88.660 & $+$1.032 & 21 02 02 & $+$48 02 06 & IRAS & 16 \\
            86 & G088.682$+$0.310 & 88.682 & $+$0.310 & 21 05 19 & $+$47 34 16 & \ldots & \ldots \\
            87 & G089.637$+$0.172 & 89.637 & $+$0.172 & 21 09 47 & $+$48 10 55 & IRAS,mm & 16 \\
            88 & G089.732$-$0.700 & 89.732 & $-$0.700 & 21 13 57 & $+$47 39 14 & IRAS & \ldots \\
            89 & G094.240$+$0.877 & 94.240 & $+$0.877 & 21 26 45 & $+$51 57 43 & IRAS & \ldots \\
            90 & G095.003$-$1.578 & 95.003 & $-$1.578 & 21 40 58 & $+$50 40 01 & IRAS,DNe & 16 \\
            91 & G095.115$-$0.570 & 95.115 & $-$0.570 & 21 37 15 & $+$51 29 46 & IRAS & 16 \\
            92 & G095.296$-$0.937 & 95.296 & $-$0.937 & 21 39 41 & $+$51 20 31 & IRAS,2MASX & 16 \\
            93 & G096.258$-$0.192 & 96.258 & $-$0.192 & 21 41 11 & $+$52 32 10 & \ldots & \ldots \\
            94 & G097.928$-$0.261 & 97.928 & $-$0.261 & 21 49 58 & $+$53 33 40 & IRAS & 16 \\
            95 & G098.320$+$1.551 & 98.320 & $+$1.551 & 21 44 03 & $+$55 12 11 & 2MASX & \ldots \\
            96 & G099.070$+$1.200 & 99.070 & $+$1.200 & 21 49 41 & $+$55 24 54 & IRAS,2MASX,Mas & 14,16 \\
            97 & G100.729$+$0.739 & 100.729 & $+$0.739 & 22 00 58 & $+$56 04 44 & \ldots & 16 \\
            98 & G100.842$+$0.636 & 100.842 & $+$0.636 & 22 02 04 & $+$56 03 54 & \ldots & \ldots \\
            99 & G101.091$-$3.037 & 101.091 & $-$3.037 & 22 18 06 & $+$53 12 19 & \ldots & \ldots \\
            100 & G103.640$+$1.087 & 103.640 & $+$1.087 & 22 16 56 & $+$58 03 00 & IRAS & 15 \\
            101 & G105.675$-$0.237 & 105.675 & $-$0.237 & 22 35 17 & $+$57 59 53 & IRAS,2MASX & 15 \\
            102 & G105.768$+$0.059 & 105.768 & $+$0.059 & 22 34 46 & $+$58 18 05 & \ldots & 15 \\
            103 & G105.834$+$0.325 & 105.834 & $+$0.325 & 22 34 11 & $+$58 33 50 & \ldots & 15 \\
            104 & G105.907$+$0.491 & 105.907 & $+$0.491 & 22 34 02 & $+$58 44 42 & IRAS & 15 \\
            105 & G106.911$+$0.647 & 106.911 & $+$0.647 & 22 40 14 & $+$59 22 27 & \ldots & 15 \\
            106 & G142.218$+$1.432 & 142.218 & $+$1.432 & 03 27 22 & $+$58 20 24 & Mas & 17 \\
            107 & G142.244$+$1.429 & 142.244 & $+$1.429 & 03 27 31 & $+$58 19 23 & IRAS,MSX,HII,2MASX,Mas & 16,17 \\
            108 & G166.237$+$0.495 & 166.237 & $+$0.495 & 05 10 16 & $+$40 39 36 & IRAS & 16 \\
            109 & G166.813$-$3.200 & 166.813 & $-$3.200 & 04 56 55 & $+$37 57 14 & IRAS,MSX,HII,smm,2MASX & 16 \\
            110 & G167.060$+$3.464 & 167.060 & $+$3.464 & 05 25 41 & $+$41 41 53 & IRAS,smm,2MASX & 16 \\
            111 & G167.267$+$3.133 & 167.267 & $+$3.133 & 05 24 50 & $+$41 20 31 & \ldots & \ldots \\
            112 & G167.414$+$3.452 & 167.414 & $+$3.452 & 05 26 41 & $+$41 23 52 & IRAS & \ldots \\
            113 & G168.122$+$3.067 & 168.122 & $+$3.067 & 05 27 04 & $+$40 35 47 & HII & \ldots \\
            114 & G168.471$-$0.972 & 168.471 & $-$0.972 & 05 11 00 & $+$37 59 28 & IRAS,HII & 16 \\
            115 & G169.838$+$1.923 & 169.838 & $+$1.923 & 05 27 04 & $+$38 32 07 & \ldots & 16 \\
            116 & G169.857$+$1.925 & 169.857 & $+$1.925 & 05 27 08 & $+$38 31 16 & IRAS,RNe & 16 \\
            117 & G169.921$+$2.052 & 169.921 & $+$2.052 & 05 27 52 & $+$38 32 17 & IRAS & 16 \\
            118 & G170.311$+$2.365 & 170.311 & $+$2.365 & 05 30 17 & $+$38 23 08 & IRAS & 16 \\
            119 & G171.054$+$1.908 & 171.054 & $+$1.908 & 05 30 25 & $+$37 30 53 & \ldots & 16 \\
            120 & G171.263$+$2.540 & 171.263 & $+$2.540 & 05 33 40 & $+$37 41 05 & IRAS & 16 \\
            121 & G171.611$+$2.349 & 171.611 & $+$2.349 & 05 33 49 & $+$37 17 19 & \ldots & \ldots \\
            122 & G171.834$+$1.644 & 171.834 & $+$1.644 & 05 31 28 & $+$36 43 06 & \ldots & \ldots \\
            123 & G171.853$+$1.637 & 171.853 & $+$1.637 & 05 31 29 & $+$36 41 55 & 2MASX & \ldots \\
            124 & G172.879$+$2.266 & 172.879 & $+$2.266 & 05 36 52 & $+$36 10 36 & IRAS,smm,2MASX,Mas & 16,17 \\
            125 & G206.529$-$3.552 & 206.529 & $-$3.552 & 06 27 04 & $+$04 03 37 & \ldots & 16 \\
            126 & G207.433$+$1.710 & 207.433 & $+$1.710 & 06 47 30 & $+$05 40 23 & IRAS,RNe & 16 \\
            127 & G207.737$+$0.964 & 207.737 & $+$0.964 & 06 45 23 & $+$05 03 47 & IRAS & 16 \\
            128 & G208.627$-$2.892 & 208.627 & $-$2.892 & 06 33 17 & $+$02 30 22 & IRAS & 16 \\
            129 & G210.415$-$1.960 & 210.415 & $-$1.960 & 06 39 52 & $+$01 20 41 & \ldots & \ldots \\
            130 & G210.469$-$2.339 & 210.469 & $-$2.339 & 06 38 37 & $+$01 07 24 & \ldots & 16 \\
            131 & G210.767$-$2.463 & 210.767 & $-$2.463 & 06 38 43 & $+$00 48 06 & \ldots & \ldots \\
            132 & G212.182$+$1.309 & 212.182 & $+$1.309 & 06 54 44 & $+$01 15 51 & IRAS & 16 \\
            133 & G217.262$-$1.467 & 217.262 & $-$1.467 & 06 54 07 & $-$04 31 22 & \ldots & \ldots \\
            134 & G218.675$-$4.420 & 218.675 & $-$4.420 & 06 46 06 & $-$07 07 08 & IRAS,RNe & 16 \\
            135 & G222.225$+$1.199 & 222.225 & $+$1.199 & 07 12 48 & $-$07 42 34 & \ldots & \ldots \\
            136 & G226.870$-$2.817 & 226.870 & $-$2.817 & 07 07 00 & $-$13 40 54 & \ldots & 16 \\
            137 & G228.098$+$0.797 & 228.098 & $+$0.797 & 07 22 31 & $-$13 05 24 & IRAS,2MASX & 16 \\

\end{longtable}
\tablefoot{
\tablefoottext{S}{Source classification from SIMBAD: IRDC stands for infrared dark cloud, of? for outflow candidate, bub for bubble, Mas for maser, (s)mm for (sub-)millimetre source, 2MASX for 2MASS extended source, RNe for reflection nebula and DNe for dark nebula.}
\tablebib{(1)~\citet{HC96}; (2) \citet{THW}; (3) \citet{HBM}; (4) \citet{BSM}; (5) \citet{WFS}; (6) \citet{SKH2002}; (7) \citet{SMC2009}; (8) \citet{BT2009}; (9) \citet{EGO}; (10) \citet{CAB2008}; (11) \citet{CF95}; (12) \citet{PBC91}; (13) \citet{CPA2006}; (14) \citet{GRC90}; (15) \citet{IRCO}; (16) \citet{WB89}; (17) \citet{Harju}; (18) \citet{SDC}.
}
}

\begin{table*}[h]
\caption{\label{newSFRs}List of location of star formation candidates that cannot be verified as clusters. Notes and references are as in Table \ref{newClusters}.}
\centering
         {\tiny
         \begin{tabular}{l*{8}{c}}
            \hline\hline
            \noalign{\smallskip}
\# & ID & $l$ & $b$ & $\alpha$\ (J2000) & $\delta$\ (J2000) & Associated sources\tablefootmark{S} & References \\
 & & [$\degr$] & [$\degr$] & [$h$\ $m$\ $s$] & [$\degr$\ ' \ ``]  &  &  \\
            \noalign{\smallskip}
            \hline
            \noalign{\smallskip}

            1 & G035.028$+$0.351 & 35.028 & $+$0.351 & 18 54 01 & $+$02 01 34 & IRAS,MSX,HII,smm,mm,Mas,of?,IRDC & 2,3,6,9,18 \\
            2 & G035.265$+$1.365 & 35.265 & $+$1.365 & 18 50 50 & $+$02 41 53 & \ldots & \ldots \\
            3 & G035.359$-$1.772 & 35.359 & $-$1.772 & 19 02 11 & $+$01 21 04 & HII & \ldots \\
            4 & G036.401$-$1.763 & 36.401 & $-$1.763 & 19 04 03 & $+$02 16 52 & IRAS,HII & \ldots \\
            5 & G037.545$-$0.111 & 37.545 & $-$0.111 & 19 00 16 & $+$04 03 14 & IRAS,MSX,HII,smm,mm,Mas,IRDC & 1,2,3,6,18 \\
            6 & G040.080$+$1.510 & 40.080 & $+$1.510 & 18 59 07 & $+$07 02 56 & \ldots & \ldots \\
            7 & G055.364$+$0.186 & 55.364 & $+$0.186 & 19 33 23 & $+$19 56 35 & IRAS,MSX,mm & \ldots \\
            8 & G059.360$-$0.206 & 59.360 & $-$0.206 & 19 43 18 & $+$23 13 59 & IRAS,MSX,HII,smm,mm & 4,5,10 \\
            9 & G059.640$-$0.181 & 59.640 & $-$0.181 & 19 43 48 & $+$23 29 17 & IRAS,MSX,HII,smm,mm,IRDC & 10,18 \\
            10 & G061.315$-$2.062 & 61.315 & $-$2.062 & 19 54 36 & $+$23 58 41 & IRAS & \ldots \\
            11 & G068.858$-$0.041 & 68.858 & $-$0.041 & 20 04 44 & $+$31 27 27 & \ldots & \ldots \\
            12 & G076.855$+$0.761 & 76.855 & $+$0.761 & 20 23 06 & $+$38 33 47 & \ldots & \ldots \\
            13 & G077.901$+$1.769 & 77.901 & $+$1.769 & 20 21 55 & $+$39 59 49 & IRAS,MSX,2MASX & \ldots \\
            14 & G078.121$+$3.632 & 78.121 & $+$3.632 & 20 14 26 & $+$41 13 28 & IRAS,MSX,HII,smm,Mas & 4,5,6,11,12,17 \\
            15 & G078.235$+$0.901 & 78.235 & $+$0.901 & 20 26 37 & $+$39 46 16 & IRAS,mm & \ldots \\
            16 & G079.151$+$1.830 & 79.151 & $+$1.830 & 20 25 25 & $+$41 03 22 & IRAS & \ldots \\
            17 & G079.155$+$2.222 & 79.155 & $+$2.222 & 20 23 44 & $+$41 17 04 & \ldots & \ldots \\
            18 & G079.482$-$0.718 & 79.482 & $-$0.718 & 20 37 15 & $+$39 49 01 & mm & \ldots \\
            19 & G079.852$-$1.507 & 79.852 & $-$1.507 & 20 41 41 & $+$39 37 48 & \ldots & \ldots \\
            20 & G081.309$-$0.112 & 81.309 & $-$0.112 & 20 40 35 & $+$41 38 13 & mm & \ldots \\
            21 & G081.516$+$0.192 & 81.516 & $+$0.192 & 20 39 58 & $+$41 59 12 & MXS,2MASX & \ldots \\
            22 & G085.034$+$0.361 & 85.034 & $+$0.361 & 20 51 19 & $+$44 50 31 & mm,DNe & \ldots \\
            23 & G162.459$-$8.674 & 162.459 & $-$8.674 & 04 21 38 & $+$37 34 44 & IRAS & \ldots \\
            24 & G171.531$+$2.445 & 171.531 & $+$2.445 & 05 34 00 & $+$37 24 28 & IRAS & \ldots \\
            25 & G173.185$+$2.356 & 173.185 & $+$2.356 & 05 38 04 & $+$35 58 00 & smm,Mas & 16,17 \\
            26 & G207.312$-$2.538 & 207.312 & $-$2.538 & 06 32 07 & $+$03 50 05 & IRAS & 16 \\
            27 & G209.561$+$0.577 & 209.561 & $+$0.577 & 06 47 20 & $+$03 15 47 & IRAS & \ldots\\
            28 & G212.064$-$0.739 & 212.064 & $-$0.739 & 06 47 13 & $+$00 26 06 & IRAS,MSX,Mas & 16,17 \\
            29 & G217.496$-$0.070 & 217.496 & $-$0.070 & 06 59 32 & $-$04 05 34 & \ldots & \ldots \\
            30 & G218.053$-$0.117 & 218.053 & $-$0.117 & 07 00 23 & $-$04 36 36 & IRAS,HII,Mas & 16,17 \\

            \noalign{\smallskip}
            \hline
\end{tabular}}
\end{table*}

\begin{table*}[ht]
\caption{\label{Pubs}List of publications referenced in Tables \ref{newClusters} and \ref{newSFRs}.}
\centering
         {\tiny
         \begin{tabular}{*{4}{c}}
            \hline\hline
            \noalign{\smallskip}
            \# & BibCode & Aut & Description \\
            \noalign{\smallskip}
            \hline
            \noalign{\smallskip}

            1 & 1996A\&AS..120..283H & Hofner \& Churchwell & Water maser emission of UC HII regions \\
            2 & 2006A\&A...453.1003T & Thompson et al. & SCUBA smm survey of IRAS and UC HII regions \\
            3 & 2005MNRAS.363..405H & Hill et al. & mm observations of SFRs \\
            4 & 2002ApJ...566..945B & Beuther et al. & SFRs in early stage of evolution - dust cores of HMPOs \\
            5 & 2004A\&A...417..115W & Williams et al. & smm survey of HMPOs \\
            6 & 2000A\&AS..143..269S & Szymczak et al. & Methanol maser emission from IRAS sources \\
            7 & 2009A\&A...504..415S & Schuller et al. & APEX smm survey of protostellar clumps \\
            8 & 2009ApJ...696..484B & Butler \& Tan & mm survey of IRDCs \\
            9 & 2008AJ....136.2391C & Cyganowski et al. & MYSO outflow candidates \\
            10 & 2008ApJ...681..428C & Chapin et al. & BLAST smm survey of HMPOs \\
            11 & 1995A\&A...302..521C & Codella \& Felli & Water masers without associated HII regions \\
            12 & 1991A\&A...246..249P & Palla et al. & Water masers associated with molecular clouds and UC HII regions \\
            13 & 2006ApJ...649..759C & Churchwell et al. & Bubble candidates from GLIMPSE \\
            14 & 1990RMxAA..20...51G & Gyulbudaghian et al. & Water masers in IRAS sources \\
            15 & 2003A\&A...399.1083K & Kerton \& Brunt & CO emission of IRAS sources \\
            16 & 1989A\&AS...80..149W & Wouterloot \& Brand & CO observations of IRAS sources with colours of SFRs \\
            17 & 1998A\&AS..132..211H & Harju et al. & SiO emission of masers \\
            18 & 2009A\&A...505..405P & Peretto \& Fuller & GLIMPSE IRDCs: initial conditions of stellar protocluster formation \\

            \noalign{\smallskip}
            \hline
\end{tabular}}
\end{table*}

\Online

\begin{appendix}
\section{False positive clusters}

\subsection{False positive clusters caused by artefacts}\label{appA1}
\begin{figure*}
\centering
\includegraphics[width=16cm]{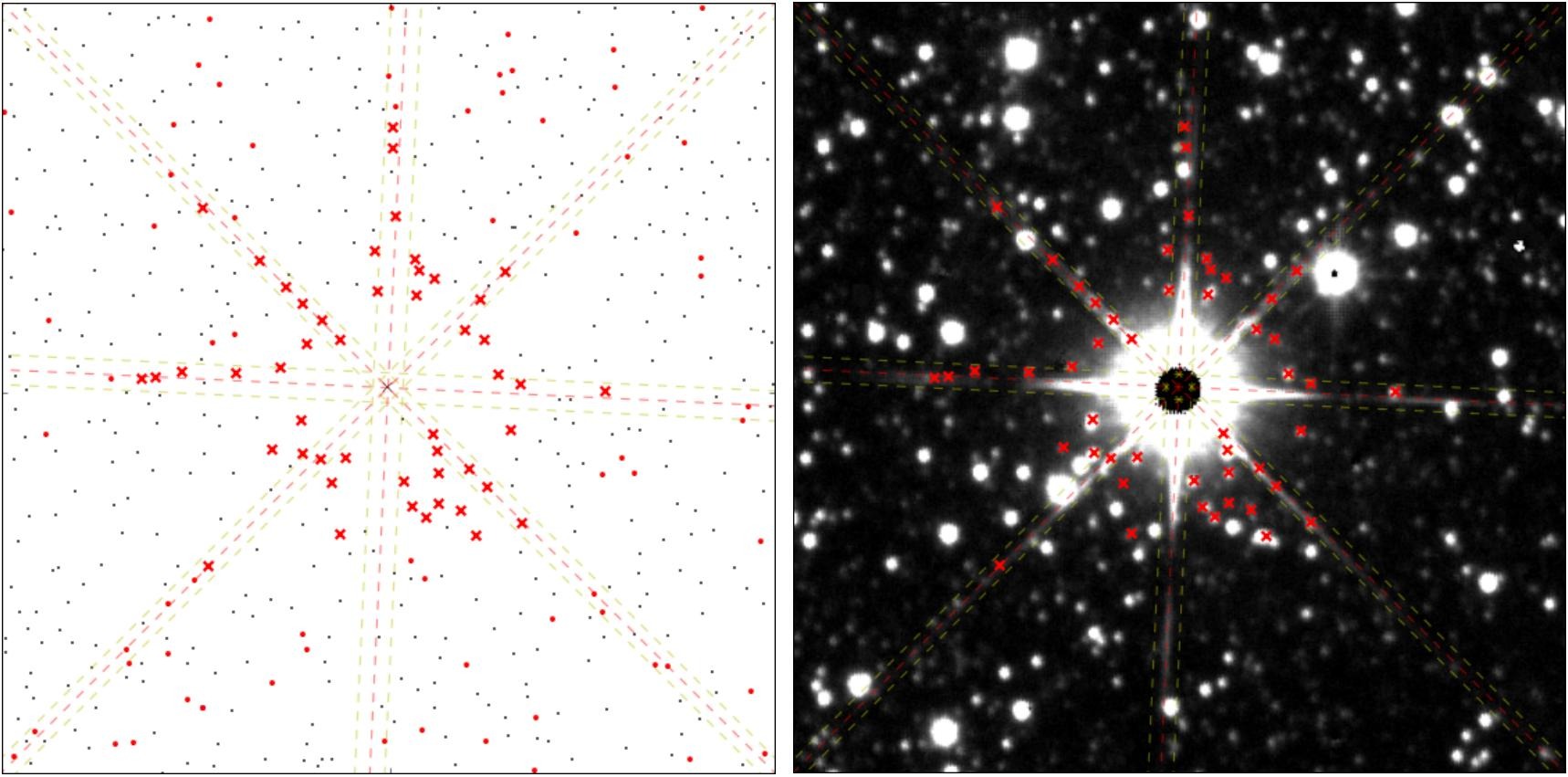}
\caption{Example of a false positive cluster caused by the bright $K=4\fm4$ star 2MASS J18360783-0644313. The red points and crosses are UKIDSS GPS sources brighter than $17^\mathrm{m}$ and classified as non-stellar. The red crosses mark the filtered out points.}
  \label{diffrSpikes}%
\end{figure*}

\begin{figure*}
\centering
\includegraphics[width=16cm]{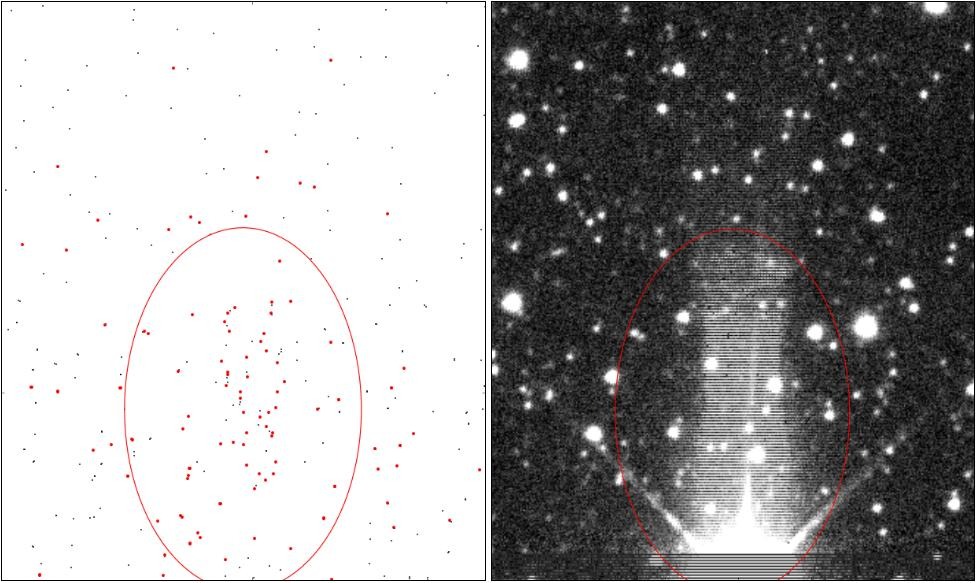}
\caption{False positive cluster caused by the $K=5\fm1$ star 2MASS J19013383+1116532 at the array edge. The red points are UKIDSS non-stellar sources brighter than $17^\mathrm{m}$ in \textit{K}.}
  \label{brightAtBorder}%
\end{figure*}

\begin{figure*}
\centering
\includegraphics[width=16cm]{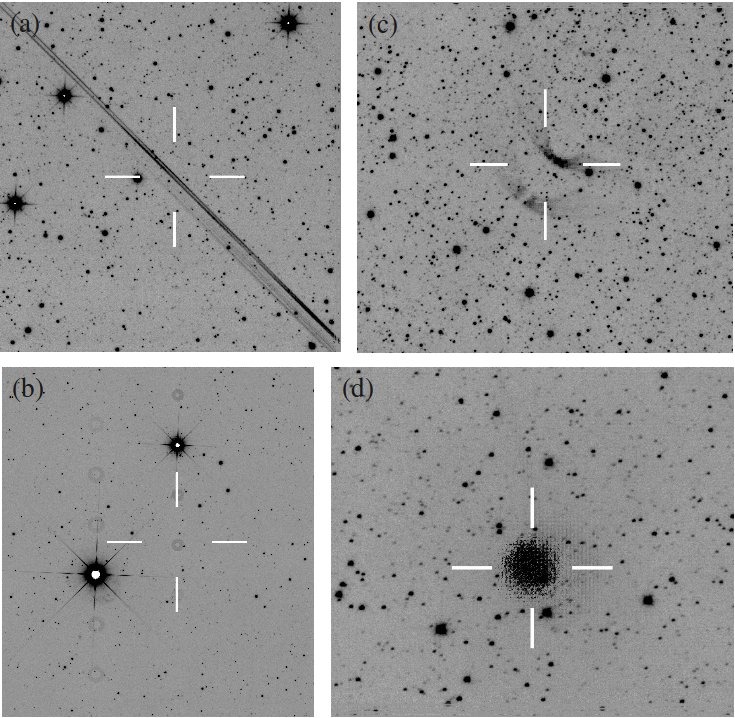}
\caption{Examples of false positive clusters due to: a) a beam, b) cross-talk images, c) 'bow-ties' (the bright star causing the artefact is 5\arcmin\ towards the upper right corner) and d) a persistence image.}\label{falsePos}%
\end{figure*}

Clustering all sources without filtering the data fails. Clustering only sources with \textit{K} magnitude brighter than $17^\mathrm{m}$, \texttt{k\_1ppErrBits} $<524288$ and \texttt{mergedClass} $=+1$ improves the results remarkably, but visual inspection of the images of the candidate areas revealed a large fraction of the cluster candidates to be blatant false positives:
   \begin{enumerate}[i)]
   \item \label{spikeHalo} Bright stars tend to create artefacts in the catalogue data appearing as \texttt{mergedClass} $=+1$ classifications.
       This happens specially in the direction of the 8 spikes
       of the diffraction pattern from the two spiders supporting the secondary and the guider auxiliary lens (see Sect. 7.6 in \citet{Dye}).
       We fetched 2MASS stars brighter than $10^\mathrm{m}$ in \textit{K}, and examined their surroundings in the UKIDSS GPS images and catalogues creating thus rules according to which non-stellar sources are discarded both
       very near the bright star, and also farther away in the direction of the 8 diffraction spikes. The brighter the star, the greater the distance to which it
       produces false classifications.
       Here we note that bits 2 and 20 for the quality error bit flag are not yet implemented.
       The issue for the former is `Close to a bright source' and for the latter `Possible diffraction spike artefact/contamination' \citep{qualityErr}.
       An example is shown in Fig. \ref{diffrSpikes}.
   \item \label{edgeEff} Bright stars at or just outside an array edge tend to create \texttt{mergedClass} $=+1$ classifications.
       To find such potential locations we compare the coordinates of 2MASS stars brighter than $8^\mathrm{m}$ in \textit{K} against parameters \texttt{minRa}, \texttt{minDec}, \texttt{maxRa}
       and \texttt{maxDec} in the UKIDSS table \texttt{CurrentAstrometry}, and check that parameter \texttt{multiframeID} equals in
       the tables \texttt{CurrentAstrometry} and \texttt{gpsDetection}. Each cluster candidate is compared to these locations in order to automatically
       remove false positives. This method might remove also true positives as is the case with e.g. \object{[BDB2003] G094.60-01.80}. An example is shown in Fig. \ref{brightAtBorder}.
   \item Beams, 'bow-ties', cross-talk images and persistence images create clusters of non-stellar sources.
       The 'bow-tie' is a low-level feature in the PSF produced by haloes of bright stars (see Sect. 7.6 in \citet{Dye}).
       As for the cross-talk images the WSA states that \it the GPS cannot ever be cross-talk flagged with the current algorithm parameters as its fields are just simply too crowded \rm \citep{qualityErr}.
       The first three types of these false positives are not numerous. It would be useful to remove the persistence image clusters but at present we cannot separate
       them from true positive clusters using the catalogue data. Examples are shown in Fig. \ref{falsePos}.
   \end{enumerate}

\subsection{False positive clusters caused by surface brightness}\label{appA2}
\begin{figure*}[htb!]
\centering
\includegraphics[width=\textwidth]{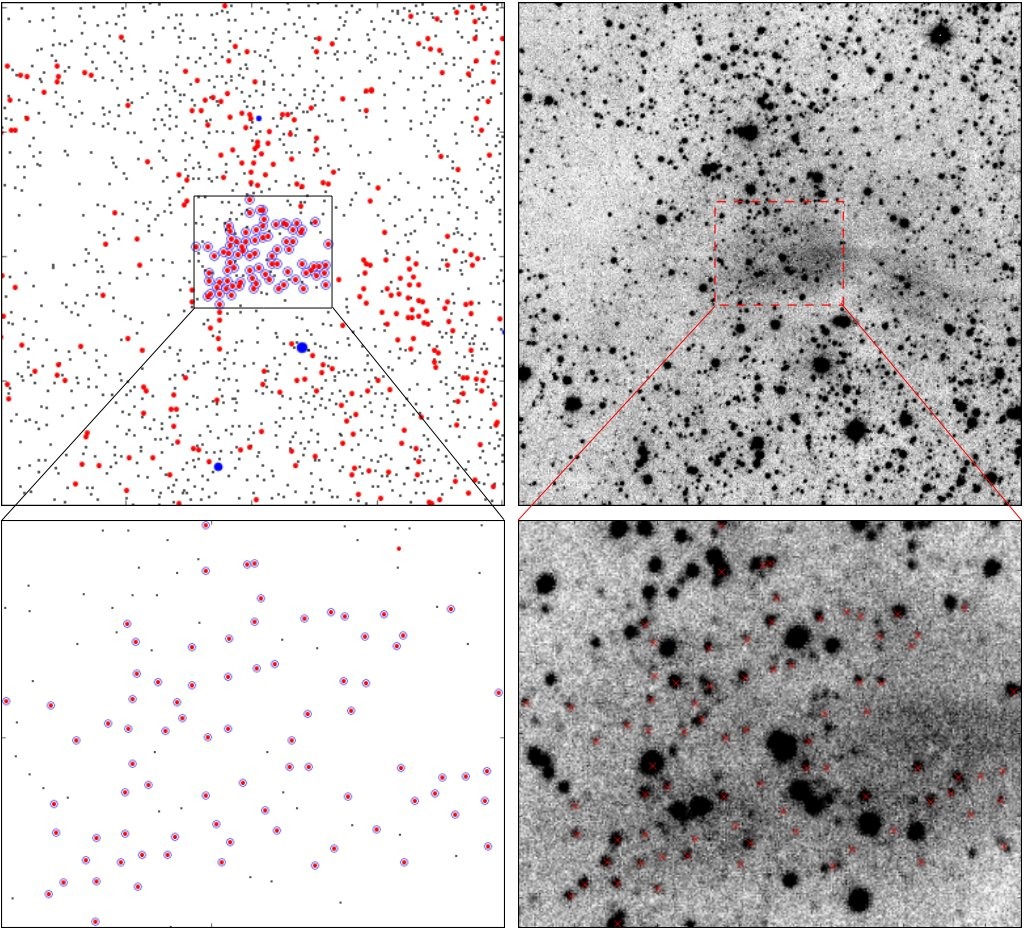}
\caption{False positive cluster at ($l=12.841\degr,b=0.544\degr$) caused by surface brightness. The panels on the left show the catalogue data and the panels on the right the corresponding on \textit{K} band images. The red points are the UKIDSS sources classified as non-stellar and brighter than $17^\mathrm{m}$ in the \textit{K} band. The blue circles mark the cluster members given by the algorithm. The blue points are sources listed in 2MASS but not in UKIDSS GPS. Image orientation is North up and East left.}
\label{falsePosEmisA}
\end{figure*}

\begin{figure*}[htb!]
\centering
\includegraphics[width=\textwidth]{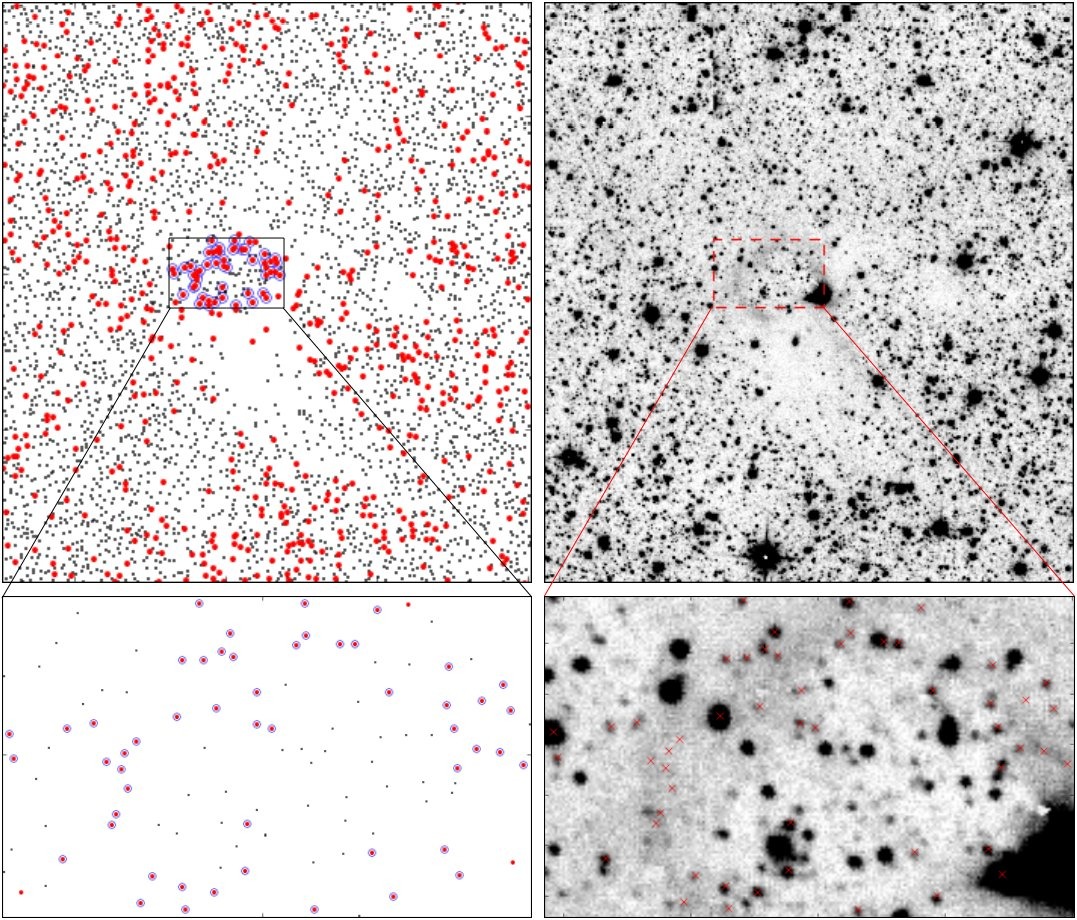}
\caption{False positive cluster at ($l=16.799\degr,b=0.126\degr$) caused by surface brightness. The catalogue data is shown on the left and the corresponding \textit{K} band image on the right. The symbols are as in Fig. \ref{falsePosEmisA}.}
\label{falsePosEmisB}
\end{figure*}

In Figs. \ref{falsePosEmisA} and \ref{falsePosEmisB} two examples of a false positive candidate caused by surface brightness are shown. In Fig. \ref{falsePosEmisA} the false positive cluster at ($l=12.841\degr,b=0.544\degr$) is caused by the interplay of extinction and the reflection of the interstellar radiation field from the dust cloud. In Fig. \ref{falsePosEmisB} the object at ($l=16.799\degr,b=0.126\degr$) at the center of the image is a millimetre radio-source and classified as a possible planetary nebula. This seems to be an outflow coming from a hole in a dark cloud. Excess surface brightness due to the bright central source makes the stars appear as non-stellar and in addition seems to produce non-existent sources.

\section{Examples of cluster candidates}\label{appB}

\begin{figure*}[htb!]
\centering
\includegraphics[width=\textwidth]{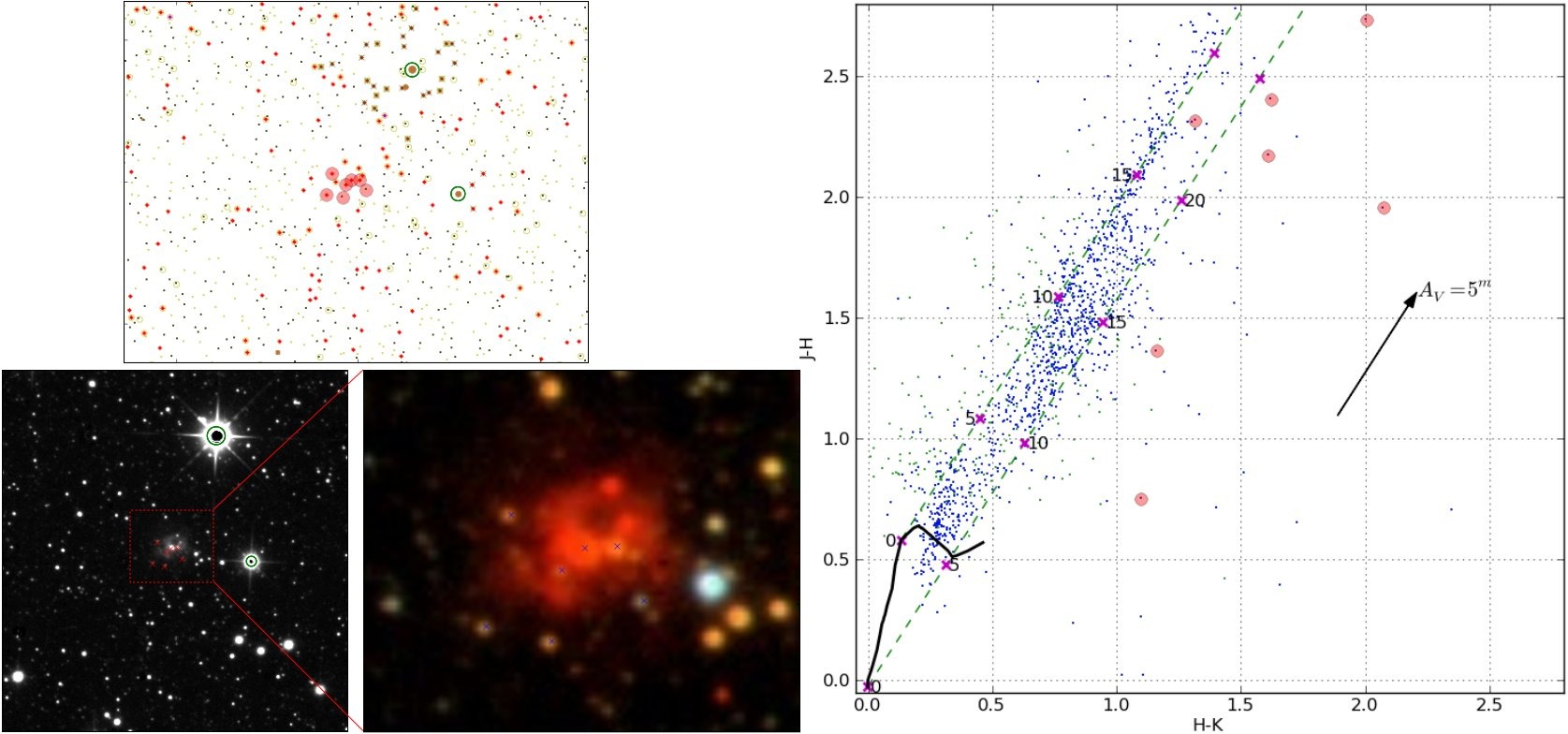}
\caption{2.5\arcmin\ by 2.5\arcmin\ box around cluster candidate 43. The red points in the panel upper left are UKIDSS non-stellar sources brighter than $17^\mathrm{m}$ in \textit{K}, black points other sources brighter than $17^\mathrm{m}$ in \textit{K} and yellow points sources fainter than $17^\mathrm{m}$ in \textit{K}. Brown points are sources listed in 2MASS but not in UKIDSS GPS. Sources encircled by a yellow line do not have coverage in all three bands. The stars encircled by a green line in the catalogue data plot and the WSA grey scale image lower left cause non-stellar classifications and produce false positive clusters: the algorithm removes the sources marked with a green cross. The WSA false colour blow-up image (\textit{J} image coded in blue, \textit{H} image in green and \textit{K} image in red) of the cluster candidate is shown to the right of the grey scale image. Image orientation is North up and East left. All the UKIDSS GPS sources within the catalogue data plot are plotted in the $(H-K, J-H)$ colour-colour plot on the right. Blue dots are sources brighter than $17^\mathrm{m}$ and green dots fainter than $17^\mathrm{m}$ in \textit{K}. The red filled circles mark UKIDSS sources (both stellar and non-stellar) in the cluster direction brighter than $17^\mathrm{m}$ in \textit{K}.}
  \label{cc43}%
\end{figure*}

\begin{figure*}[htb!]
\centering
\includegraphics[width=\textwidth]{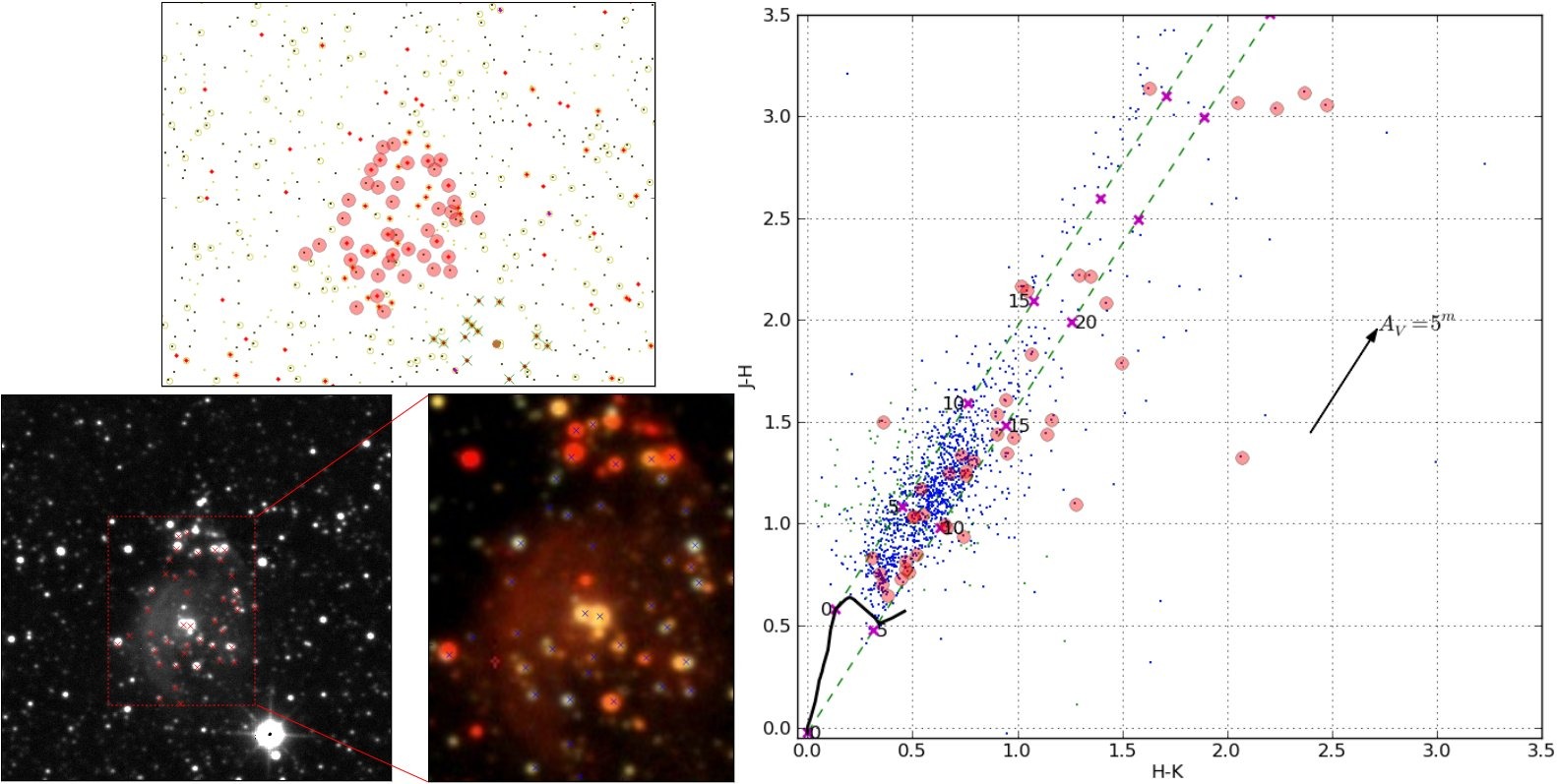}
\caption{As Fig. \ref{cc43} for cluster candidate 20. The box size is 1.9\arcmin\ by 1.9\arcmin.}
  \label{cc20}%
\end{figure*}

\begin{figure*}[htb!]
\centering
\includegraphics[width=\textwidth]{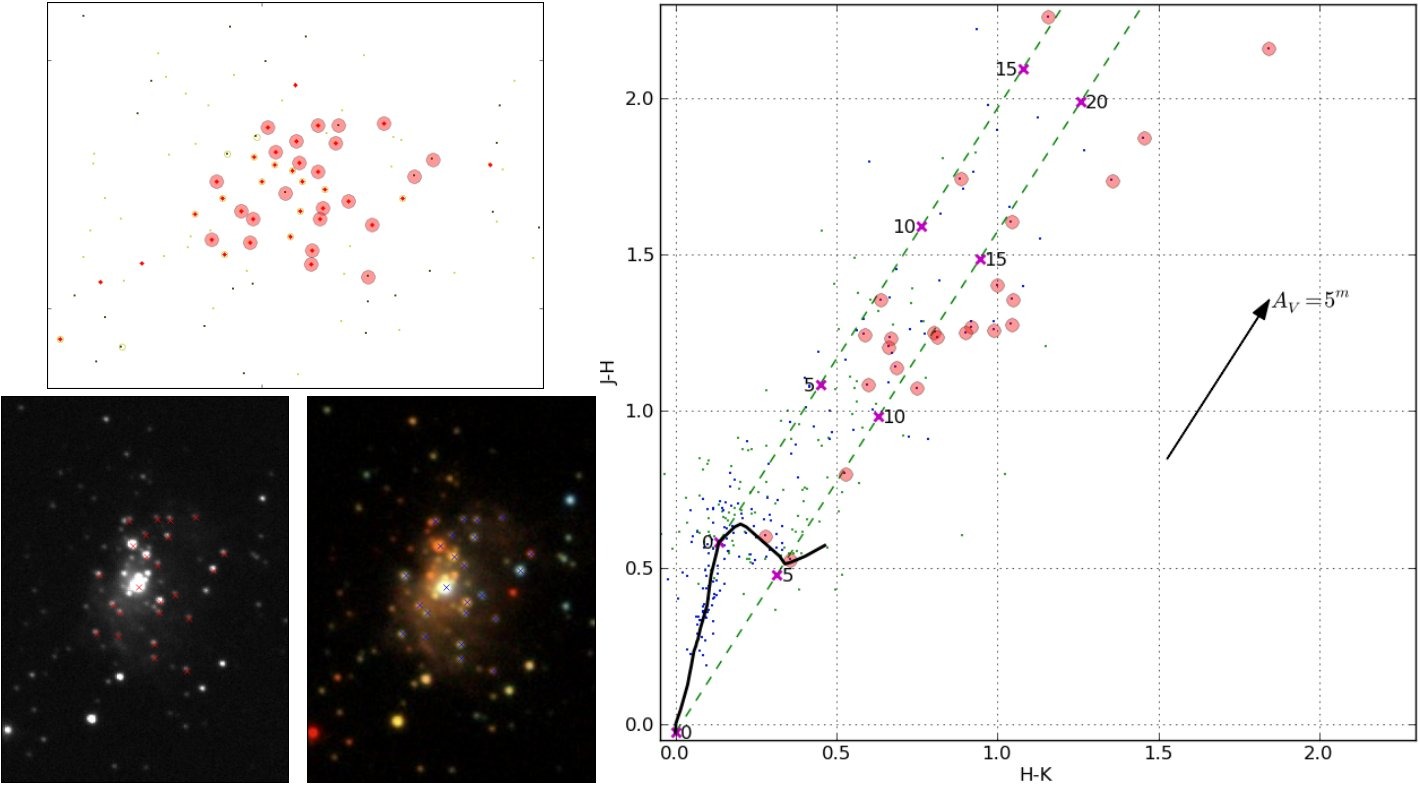}
\caption{As Fig. \ref{cc43} for cluster candidate 110. The box size is 1.2\arcmin\ by 1.2\arcmin.}
  \label{cc110}%
\end{figure*}

\begin{figure*}[htb!]
\centering
\includegraphics[width=\textwidth]{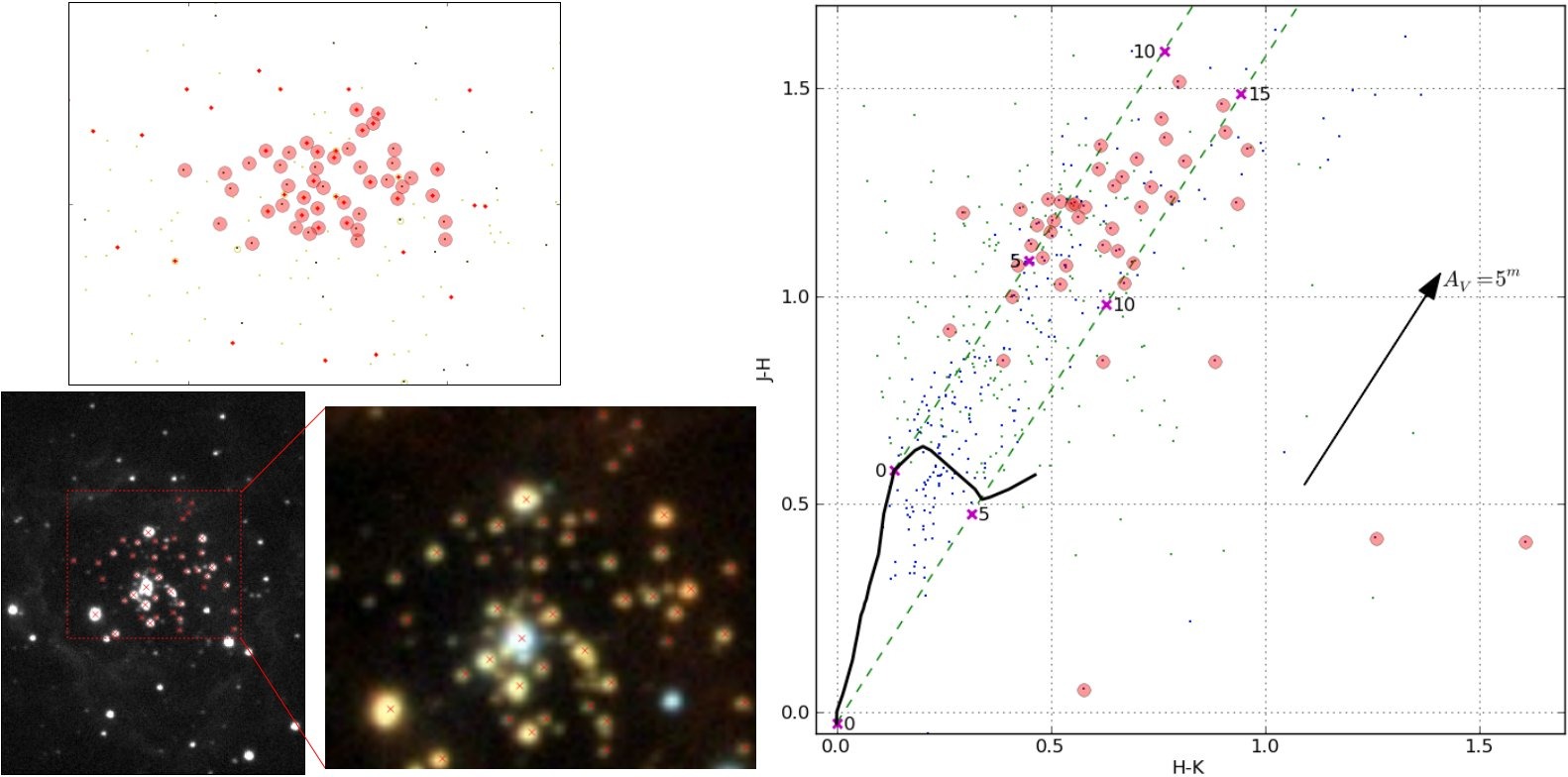}
\caption{As Fig. \ref{cc43} for cluster candidate 116. The box size is 1.4\arcmin\ by 1.4\arcmin.}
  \label{cc116}%
\end{figure*}

\begin{figure*}[htb!]
\centering
\includegraphics[width=\textwidth]{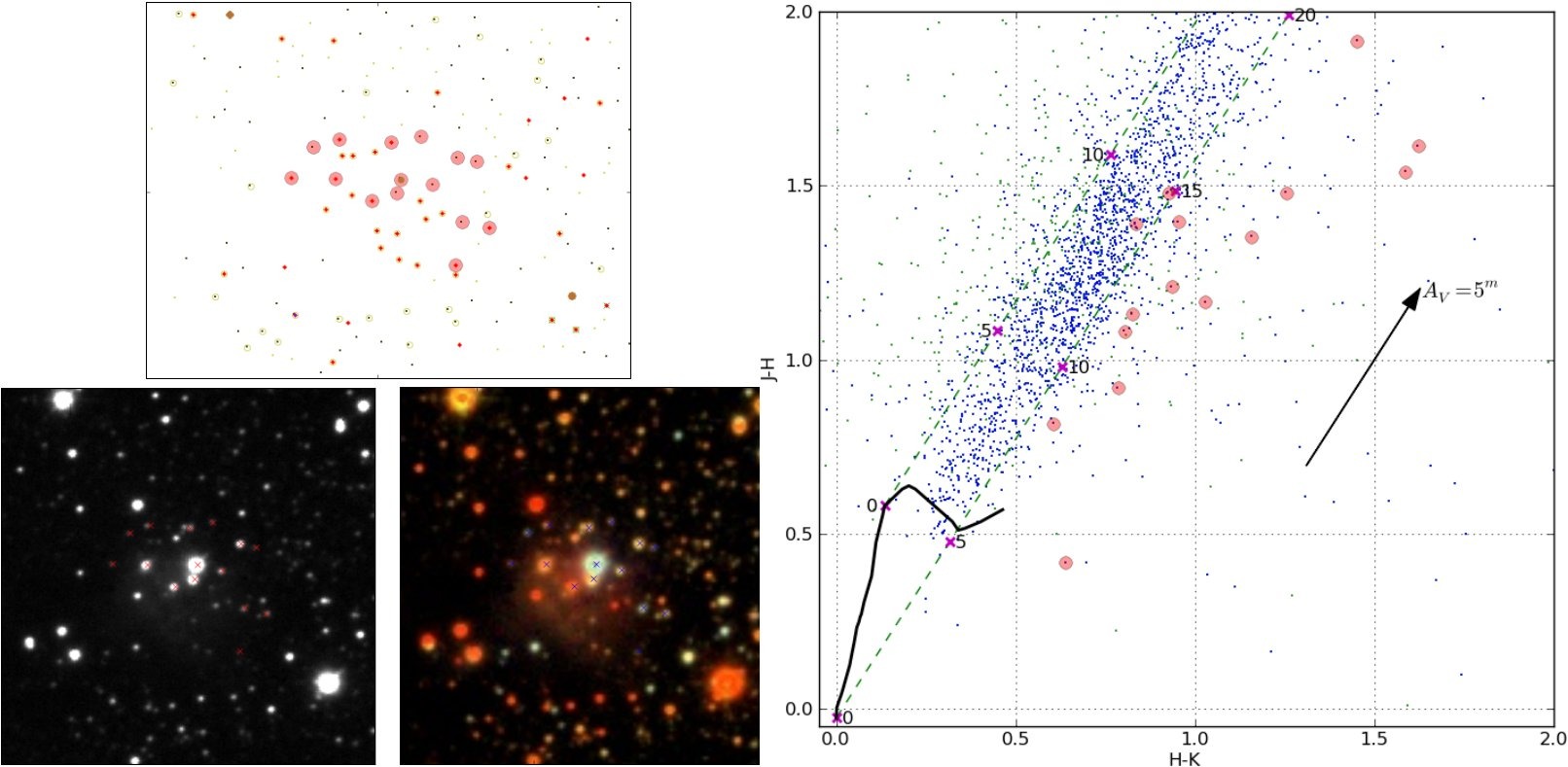}
\caption{As Fig. \ref{cc43} for cluster candidate 9. The box size is 0.9\arcmin\ by 0.9\arcmin.}
  \label{cc9}%
\end{figure*}

\begin{figure*}[htb!]
\centering
\includegraphics[width=\textwidth]{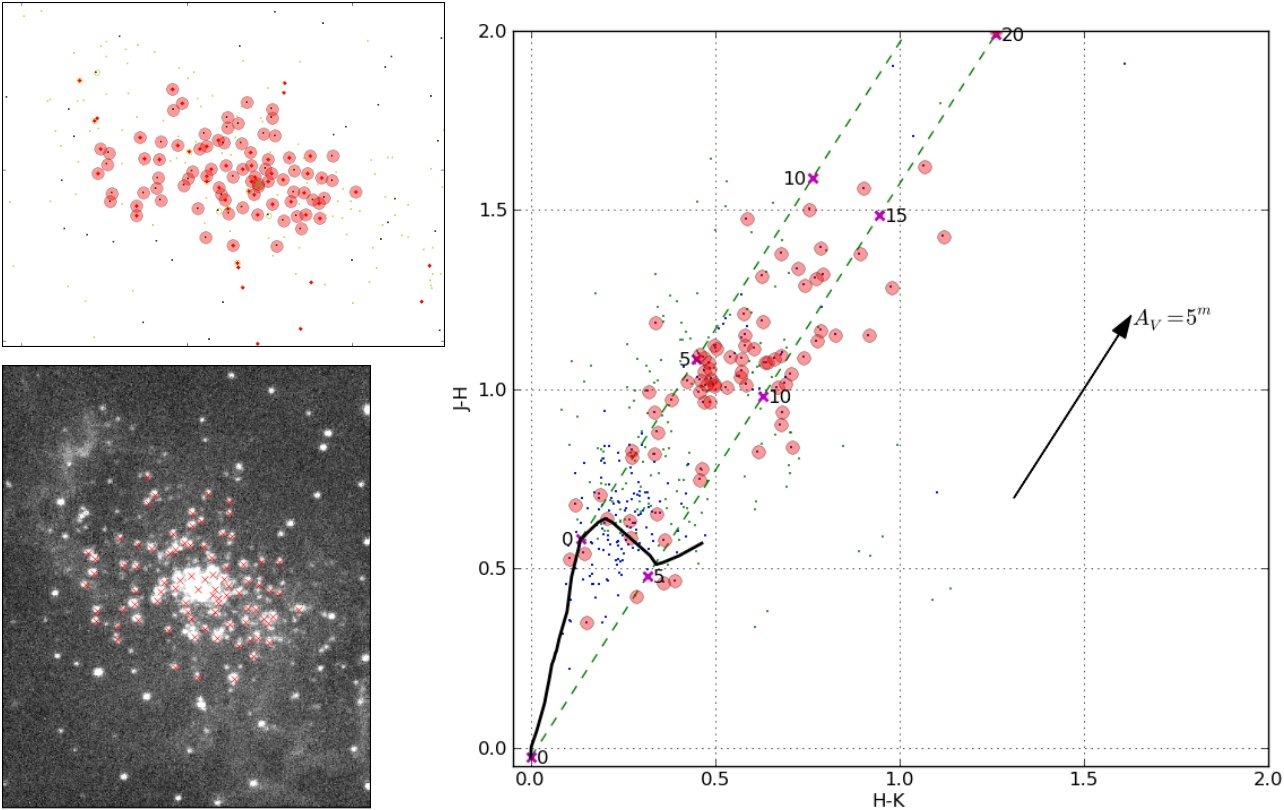}
\caption{As Fig. \ref{cc43} for cluster candidate 114. The box size is 1.8\arcmin\ by 1.8\arcmin.}
  \label{cc114}%
\end{figure*}

Example cluster candidates are shown in Figs. \ref{cc43}$-$\ref{cc114}. The different panels in the figures are as follows. Upper left: The catalogue data in the direction of the cluster candidate. The red points are UKIDSS non-stellar sources brighter than 17 magnitudes, black points other sources brighter than $17^\mathrm{m}$, and yellow points sources fainter than $17^\mathrm{m}$ in the \textit{K} filter. Brown points are stars listed in 2MASS but not in UKIDSS GPS. Sources encircled by a yellow line do not have coverage in all three bands. The bright stars encircled by a green line in the catalogue data plot and the grey scale image lower left cause non-stellar classifications and produce false positive clusters: the algorithm removes the sources indicated with a green cross. Two bright stars encircled by a green line in the catalogue data plot and the WSA grey scale plot in the lower left. The WSA false colour blow-up image (\textit{J} image coded in blue, \textit{H} image in green and \textit{K} image in red) of the cluster candidate is shown to the right of the grey scale image. All the UKIDSS GPS sources within the catalogue data plot are plotted in the $(H-K, J-H)$ colour-colour plot on the right. Blue dots mark sources brighter than $17^\mathrm{m}$ and green dots sources fainter than $17^\mathrm{m}$ in \textit{K}. The red filled circles mark UKIDSS sources (both stellar and non-stellar) in the cluster direction brighter than $17^\mathrm{m}$ in \textit{K}.
The approximate unreddened main sequence is plotted with a continuous line. Approximate main sequence reddening lines are shown with dashed lines. The numbers on the reddening lines show the optical extinction in case the star originates from the early or late main sequence. The arrow indicates an optical extinction of 5 magnitudes.

The automated search uses by default the \textit{AperMag3} magnitudes (2.0\arcsec\ aperture diameter). For the colour-colour plots we experimented also with the \textit{AperMag1} (1.0\arcsec\ aperture diameter) and \textit{AperMag4} (2.8\arcsec\ aperture diameter) extended source magnitudes. For cluster candidates 20, 114 and 116 the colour-colour plots use the \textit{AperMag1} magnitudes because they seem to give better precision. For the remaining cluster candidates (9, 43 and 110) the colour-colour plots use the \textit{AperMag3} magnitudes. For \textit{ppErrbits} we apply the same $<524288$ limit as in the automated search knowing that by using this limit we don't take advantage of all the photometric warning flags. However for these six cluster candidates only for a negligible portion of the data \textit{ppErrbits} $>255$.

We use in the figures a reddening slope of 1.6. We recognise that reddening bands in colour-colour diagrams are delimited by curves rather than vectors (e.g. \citet{Golay} and \citet{SteadHoare}). The value of 1.6 is the mean of all the reddening tracks in \citet{SteadHoare}. Irrespective of the uncertainty of the reddening vector the colour-colour plots allow to estimate the reddening in the direction of the cluster candidates. The notable difference between cluster candidates is the larger number of field stars, especially giants, in the direction of the inner Galaxy (cluster candidates 9, 20 and 43) with respect to the number of stars in the outer Galaxy (cluster candidates 110, 114 and 116). In the inner Galaxy the field stars, i.e. the stars not classified as non-stellar, lie within the approximate reddening path outlined by the reddening lines. In the outer Galaxy the statistics is poor because of the small number of field stars and the high extinction. The spread of the sources in the direction of the cluster candidates and classified as non-stellar is much higher than for the field stars. The photometry of these sources suffers from the faintness of the stars and the high background surface brightness. However, these were the objects used to locate the cluster candidates. In general it can be noted that the extinction towards these sources is on the average higher than for the general field stars population. The extinctions range from 10 magnitudes up to 20 magnitudes and more. Dedicated observations are needed for detailed analysis of the cluster candidates.

\section{Zone of avoidance galaxies}\label{appC}
\begin{figure}
\reflectbox{
\includegraphics[width=8.8cm,angle=90]{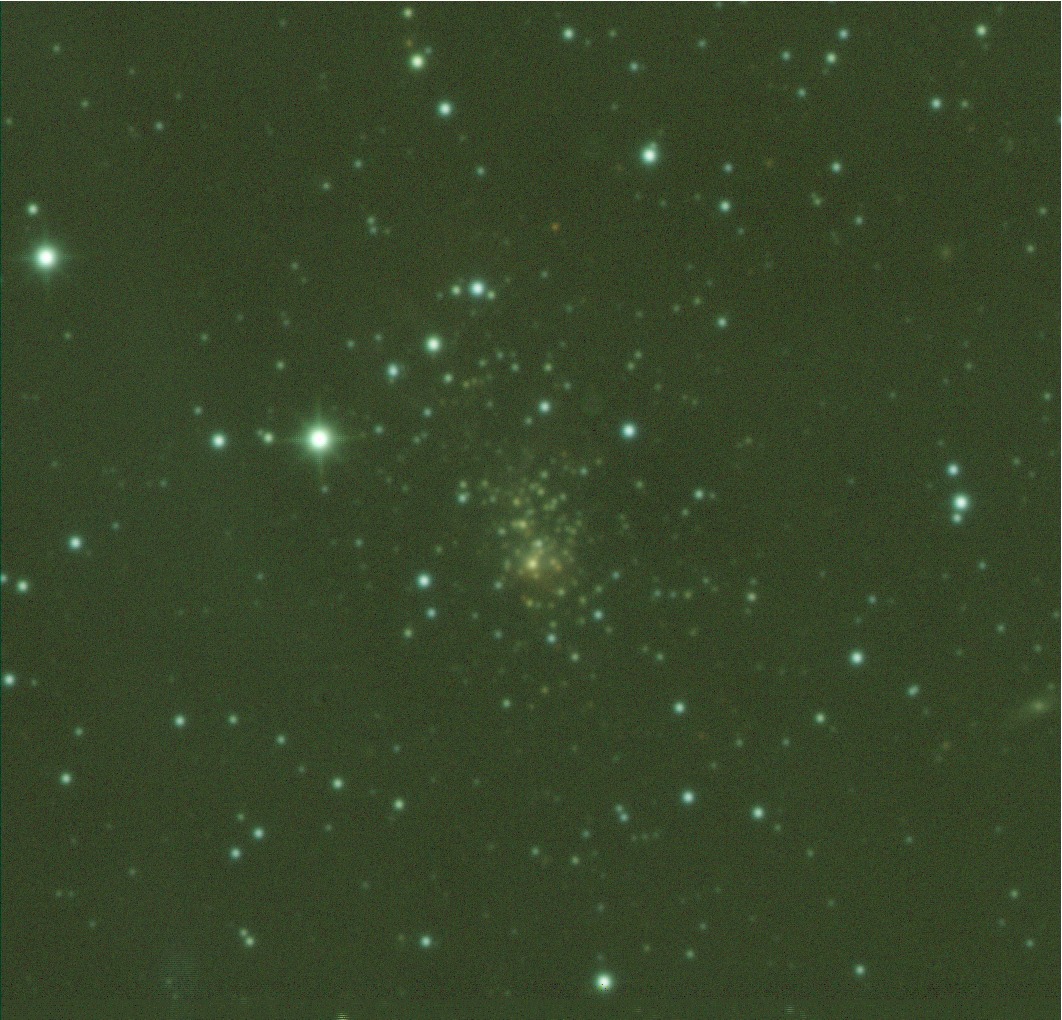}}
\caption{False colour image of cluster candidate 109. The \textit{J}, \textit{H} and \textit{K} bands are coded in blue, green and red, respectively. North up and East left. The box size is 4\arcmin\ by 4\arcmin.}
 \label{appC109}%
\end{figure}

\begin{figure}
\reflectbox{
\includegraphics[width=8.8cm,angle=180]{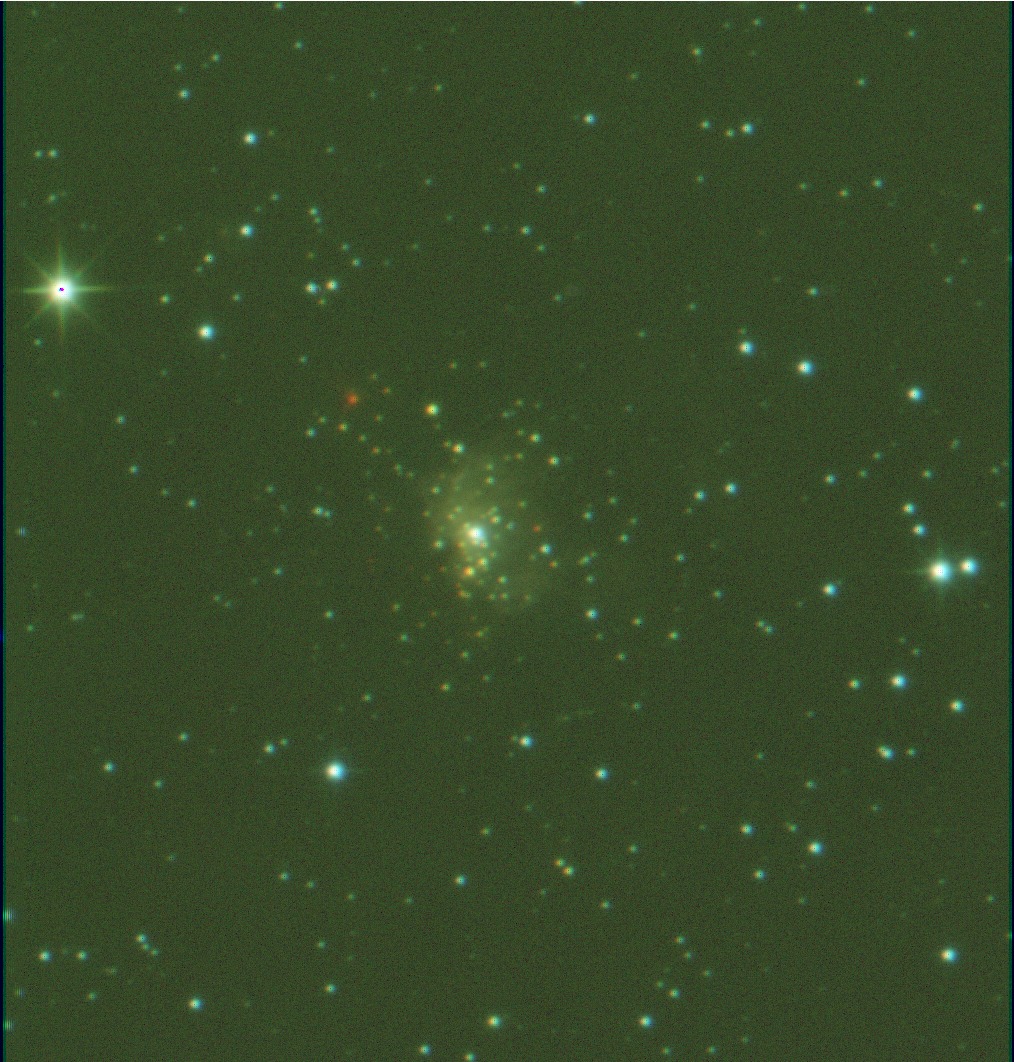}}
\caption{As Fig. \ref{appC109} for cluster candidate 110. The box size is 4\arcmin\ by 4\arcmin.}
 \label{appC110}%
\end{figure}

\begin{figure}
\reflectbox{
\includegraphics[width=8.8cm,angle=90]{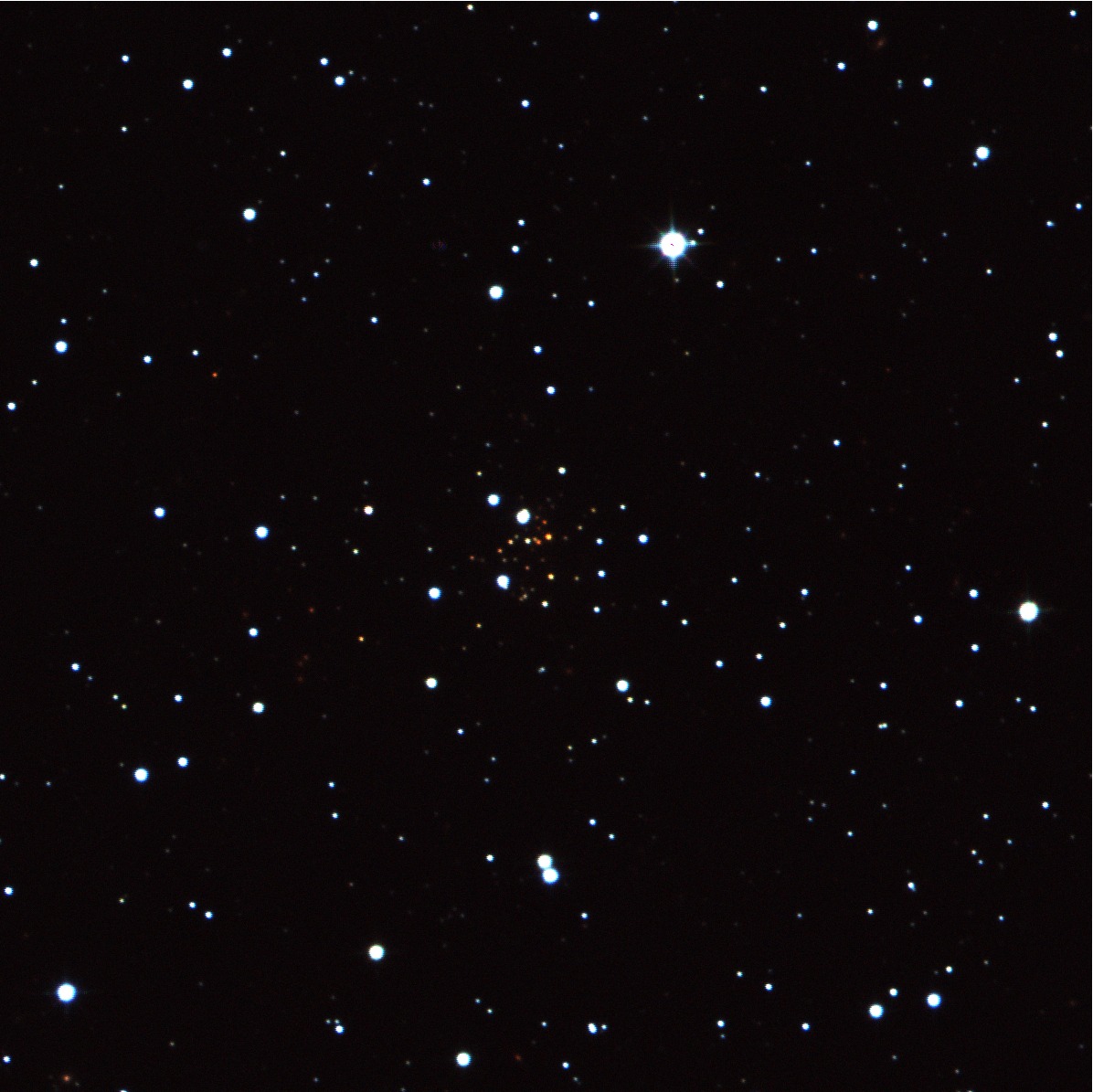}}
\caption{As Fig. \ref{appC109} for cluster candidate 112. The box size is 4\arcmin\ by 4\arcmin.}
 \label{appC112}%
\end{figure}

\begin{figure}
\reflectbox{
\includegraphics[width=8.8cm,angle=180]{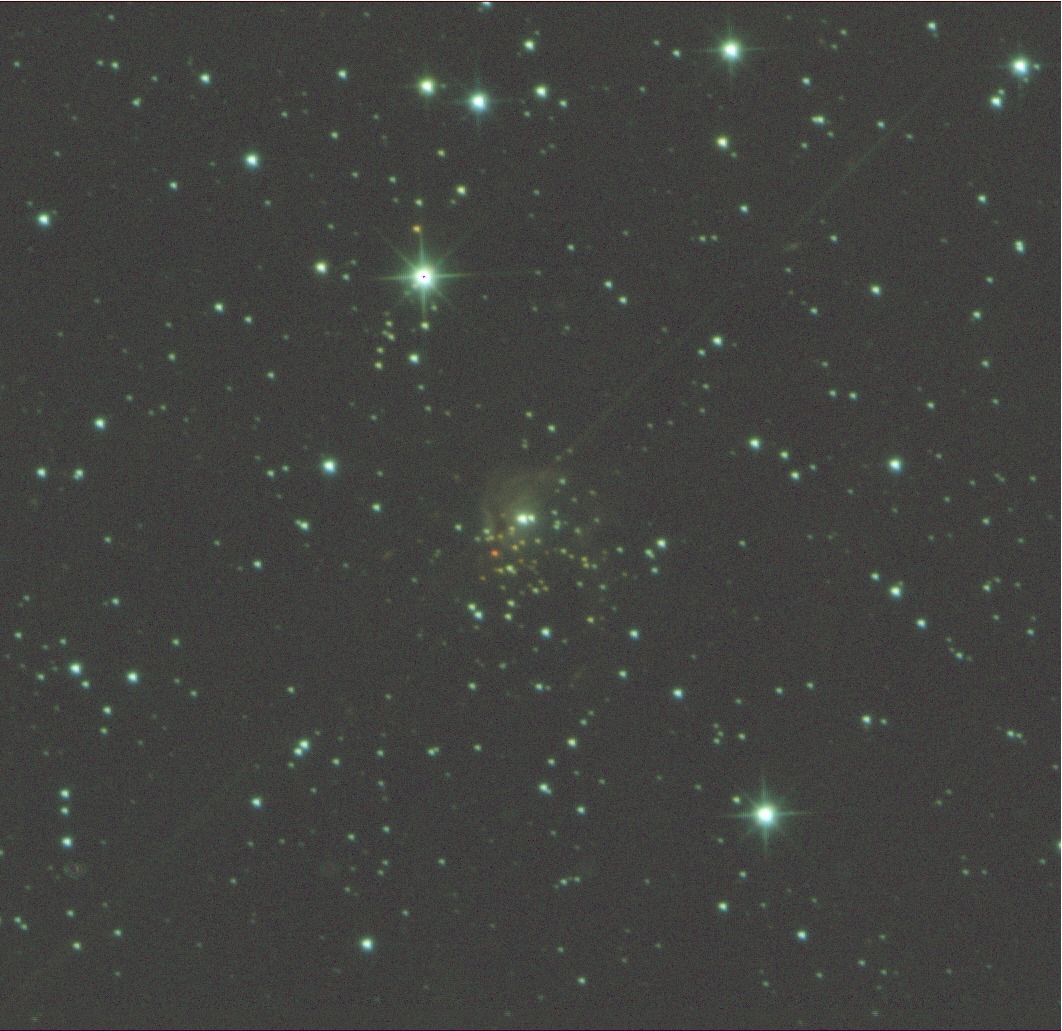}}
\caption{As Fig. \ref{appC109} for cluster candidate 137. The box size is 4\arcmin\ by 4\arcmin.}
 \label{appC137}%
\end{figure}

A zone of avoidance galaxy has been reported in the direction of three of the new cluster candidates: 109, 110, 112 and 137. False colour images of these candidates produced from WSA fits files are shown in Figs. \ref{appC109}, \ref{appC110}, \ref{appC112} and \ref{appC137}. \textit{J} image is coded in blue, \textit{H} in green and \textit{K} in red. North up and East left. A cluster of individual stars can be seen in the direction of these three cluster candidates. This would not be the case if the sources were extragalactic.

\end{appendix}

\end{document}